\documentclass[journal=jacsat,manuscript=article]{achemso}
\usepackage[version=3]{mhchem} 
\SectionNumbersOn
\usepackage{siunitx}[=v2]
\usepackage{physics} 
\usepackage{chemformula}
\usepackage{textgreek}
\usepackage{float}

\usepackage{verbatim}
\usepackage{xcolor}

\DeclareSIUnit\molar{M}
\DeclareSIUnit\photon{photon}
\DeclareSIUnit\OD{OD}

\usepackage{booktabs}
\usepackage{tabularx}
\usepackage{mciteplus}
\usepackage{makecell}

\usepackage{xcolor} 

\usepackage{soul} 

\usepackage{textgreek}

\usepackage{graphicx}			
\DeclareGraphicsExtensions{.pdf,.png,.jpg} 

\usepackage{siunitx}			
\DeclareSIUnit\photon{photon}
\DeclareSIUnit\molar{M}
\DeclareSIUnit\Molar{M}
\DeclareSIUnit\rev{rev} 
\DeclareSIUnit\OD{OD}
\DeclareSIUnit\cal{cal} 
\DeclareSIUnit\calorie{cal} 
\DeclareSIUnit\lines{lines}

\usepackage{bm} 

\DeclareUnicodeCharacter{2032}{'} 
\DeclareUnicodeCharacter{2212}{--} 

\author{Robert H. Gordon}
\affiliation{School of Mathematical and Physical Sciences, University of Sheffield, Sheffield S3 7RH, UK}
\email{rob.howard.gordon@gmail.com}

\author{Rachel C. Kilbride}
\affiliation{Department of Physics, University of Warwick, Coventry, UK}
\alsoaffiliation{XMaS, The UK X-ray Materials Science Facility, European Synchrotron Radiation Facility, Grenoble, France}

\author{Michael Hutton}
\affiliation{Department of Materials, The University of Manchester, Manchester, UK}

\author{Jenny Clark}
\affiliation{School of Mathematical and Physical Sciences, University of Sheffield, Sheffield S3 7RH, UK}

\author{Robert D.J Oliver}
\affiliation{School of Chemical, Materials and Biological Engineering, University of Sheffield, Sir Robert Hadfield Building, Mappin Street, Sheffield, S1 3JD, UK}
\email{robert.oliver@sheffield.ac.uk}

\title[An \textsf{achemso} demo]
  {Dual-Resonator Organic Microcavities for Triplet--Triplet Annihilation Upconversion}

\abbreviations{%
  3D: three-dimensional;
  GIWAXS: grazing-incidence wide-angle X-ray scattering;
  ISC: intersystem crossing;
  LDOS: local density of optical states;
  LPB: lower polariton branch;
  MINERVA: Microstructure and Interface Evolution in Realtime under VAcuum;
  NIR: near-infrared;
  PdPC: palladium phthalocyanine;
  PL: photoluminescence;
  SF: singlet fission;
  TET: triplet energy transfer;
  TMDs: transition metal dichalcogenides;
  TMM: transfer-matrix method;
  TTA: triplet--triplet annihilation;
  TTA-UC: triplet--triplet annihilation upconversion;
  UCPL: upconversion photoluminescence%
}

\begin{document}

\begin{tocentry}





\includegraphics[scale=1.0]{"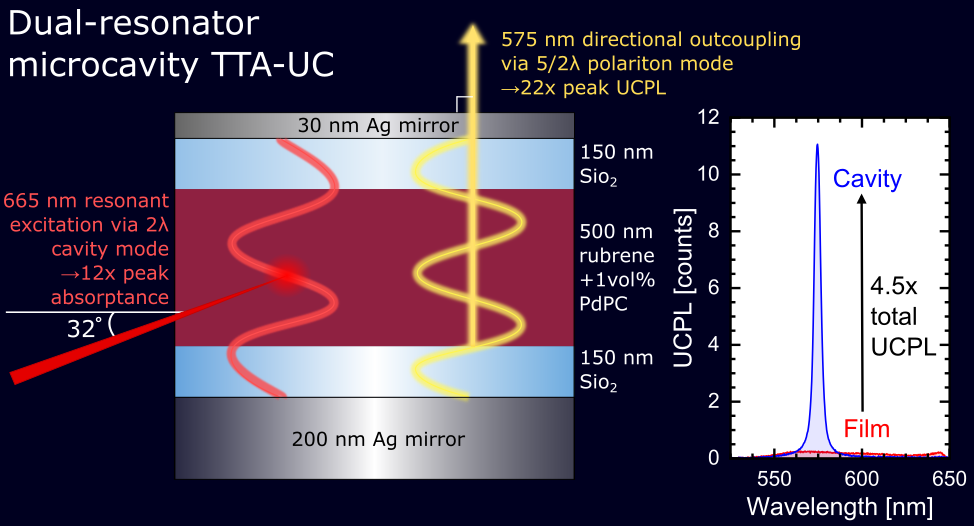"}

\end{tocentry}




\begin{abstract}\label{abstract}

Solid-state triplet--triplet annihilation upconversion offers a compelling route to harvest sub-bandgap photons to increase the efficiency of photovoltaics, but low sensitizer doping required to suppress non-radiative aggregation quenching severely restricts light absorption. Here, we present a dual-resonator planar silver microcavity housing a thermally co-evaporated rubrene:palladium phthalocyanine (PdPC) active layer designed to simultaneously enhance triplet-sensitizer absorption and out-couple upconverted emission. By aligning a dispersive cavity mode with the PdPC absorption peak at 32°, we achieve a 12-fold enhancement in absorptance at 665 nm. This absorption boost, combined with directional outcoupling through a second cavity mode, outweighs cavity-induced reabsorption losses to yield a 22-fold peak and 4.5-fold integrated enhancement in detected upconversion emission. Overall, this work highlights the potential of microcavity integration to substantially enhance solid-state upconversion performance across low-intensity optoelectronic applications.

\centering
\vspace{6pt}
\includegraphics[scale=1.0]{"figures/PdPC_paper/graphical_abstract.png"}

\end{abstract}


Optical microcavities are a proven tool for enhancing the emissive properties of organic semiconducting materials. By designing the microcavity to resonate with the emitter, the rate of spontaneous emission can be improved.\cite{purcellSpontaneousEmissionProbabilities1995} Combining this effect with the microcavities ability to redistribute emission in the forward direction,\cite{jordanEfficiencyEnhancementMicrocavity1996, benistyImpactPlanarMicrocavity1998} and to potentially enable greater triplet-harvesting through polariton formation,\cite{polakManipulatingMoleculesStrong2020, ye2021} makes this architecture an attractive tool for improving triplet--triplet annihilation upconversion (TTA-UC) materials.

TTA-UC offers the ability to convert two low-energy photons to a higher-energy one, which has applications including bio-sensing,\cite{wangMolecularNearinfraredTriplettriplet2024} photo-catalysis,\cite{bucchieriTripletTripletAnnihilation2025, huangTripletTripletAnnihilation2024} and biotherapy.\cite{sasakiNearInfraredOptogeneticGenome2019} As TTA-UC has the unique advantage of being functional under low-intensity incoherent light ($< 100\text{ mW cm}^{-2}$, equivalent to natural sunlight),\cite{singh-rachfordPhotonUpconversionBased2010} it is particularly appealing to use upconversion layers in-tandem with photovoltaic cells to improve sub-band gap light harvesting.\cite{kinoshitaPhotonUpconvertingSolid2020a, wiegholdDisorderBeneficialPerovskiteSensitized2020a, shengTremendouslyEnhancedPhotocurrent2021, tayebjeeShockleyqueisserMolecularApproaches2015, carrodRecentAdvancesTriplet2022, goldschmidtUpconversionPhotovoltaicsReview2015, ekins-daukesMolecularApproachIntermediate2008} By placing a microcavity that hosts a TTA-UC active layer underneath a solar cell, its reflective mirrors would not impede solar flux, but instead enhance the up-converted emission towards the cell and more efficiently harness a wider range of the solar spectrum.

Typically, TTA-UC is activated when a low-energy photon (e.g. near-infrared, NIR) is absorbed by a sensitizer molecule to generate a triplet exciton ($T_{1}$) through pathways such as inter-system crossing (ISC) or charge transfer. Subsequently, this triplet exciton is transferred from the sensitizer to an annihilator molecule: rubrene, for example, is commonly used as an annihilator due to its near-ideal energetic alignment ($2E_{T1} \approx E_{S1}$)\cite{herkstroeterTripletStateEnergies1981, yongEntangledTripletPair2017} and favorable spin statistics.\cite{bossanyiSpinStatisticsTriplet2021} Upon the annihilation of two triplets, an overall higher energy singlet state ($S_{1}$) is generated, which can then release its energy radiatively. The upconversion yield ($\Phi_{UC}$), defined as the ratio of the number of up-converted photons to the number of absorbed photons,\cite{jonesUpconversionQuantumYield2021} serves as a universal metric for comparing the efficiency of these materials.

A key challenge for microcavity-enhanced TTA-UC is finding a sensitizer that is compatible with the microcavity design. Although inorganic semiconductors, such as nanocrystals (quantum dots), offer widely tunable NIR absorption, their continuous absorption band overlaps with the emitter, causing self-absorption losses.\cite{narayananAlleviatingParasiticBack2025, xiaAchievingSpintripletExciton2020} Given we are aiming for a resonant enhancement of emission, these self-absorption losses would only be amplified in the microcavity. Similarly, 2D semiconductors like transition metal dichalcogenides (TMDs) have recently shown promise with good upconversion yields ($\Phi_{UC} = 1.2\%$),\cite{duanEfficientSolidstateInfraredtovisible2022a} but their performance hinges on a highly sensitive fabrication process, and their band-to-band transitions would again lead to greater self-absorption.

This leads us to organic-based triplet sensitizers with finite absorption peaks, such as palladium phthalocyanine (PdPC), whose upconversion properties have been extensively studied,\cite{sawaEnhancingNIRtovisiblePhoton2023a, radiunasCNTuningPathwaySuppress2025, radiunasNIRtovisPhotonUpconversion2021a, lissauNearInfraredVisiblePhoton2021a, wangHighlyefficientRedtoyellowGreen2021a, radiunasImpactTbutylSubstitution2020a, radiunasTripletSingletExciton2022a, singh-rachfordPdIIPhthalocyanineSensitizedTripletTriplet2008, radiunasEfficientNIRtovisPhoton2022a} with yields recently reaching $\Phi_{UC} = 3.0\%$ \cite{radiunasCNTuningPathwaySuppress2025}. Importantly for us, its low molecular weight allows it to be thermally evaporated and incorporated directly into rubrene films via co-deposition.

However, in order to avoid detrimental aggregation effects and back energy-transfer,\cite{torrePhthalocyaninesOldDyes2007, radiunasUnderstandingLimitationsNIRtovisible2020a} PdPC must be doped at low concentrations ($\approx$ 0.1 wt\%),\cite{radiunasCNTuningPathwaySuppress2025} severely reducing its absorption and upconversion signal. Fortunately, microcavities can resonantly enhance absorption, which has recently led to nanocrystal-based upconversion emission and intensity thresholds being improved by more than two orders of magnitude.\cite{wuNanocrystalSensitizedInfraredtoVisibleUpconversion2021} Taking inspiration from this, we implement silver-based microcavities designed to support both a cavity (${2}\lambda$) and polariton ($\frac{5}{2}\lambda$) mode that can resonantly enhance both the absorption of PdPC and emission of rubrene, respectively.


We begin by introducing the rubrene-PdPC system. While PdPC-sensitised upconversion is well-documented in literature,\cite{sawaEnhancingNIRtovisiblePhoton2023a, radiunasCNTuningPathwaySuppress2025, radiunasNIRtovisPhotonUpconversion2021a, lissauNearInfraredVisiblePhoton2021a, wangHighlyefficientRedtoyellowGreen2021a, radiunasImpactTbutylSubstitution2020a, radiunasTripletSingletExciton2022a, singh-rachfordPdIIPhthalocyanineSensitizedTripletTriplet2008, radiunasEfficientNIRtovisPhoton2022a} these studies typically employ solution-processing techniques such as spin-coating to achieve low doping concentrations to prevent detrimental aggregation effects.\cite{torrePhthalocyaninesOldDyes2007} In this work, we use vacuum co-deposition to produce the organic layers, as this approach ensures a high degree of thickness control and surface planarity required for precisely positioned and well-defined cavity modes.

\begin{figure}[H]
    \centering
    \includegraphics[width=1\linewidth]{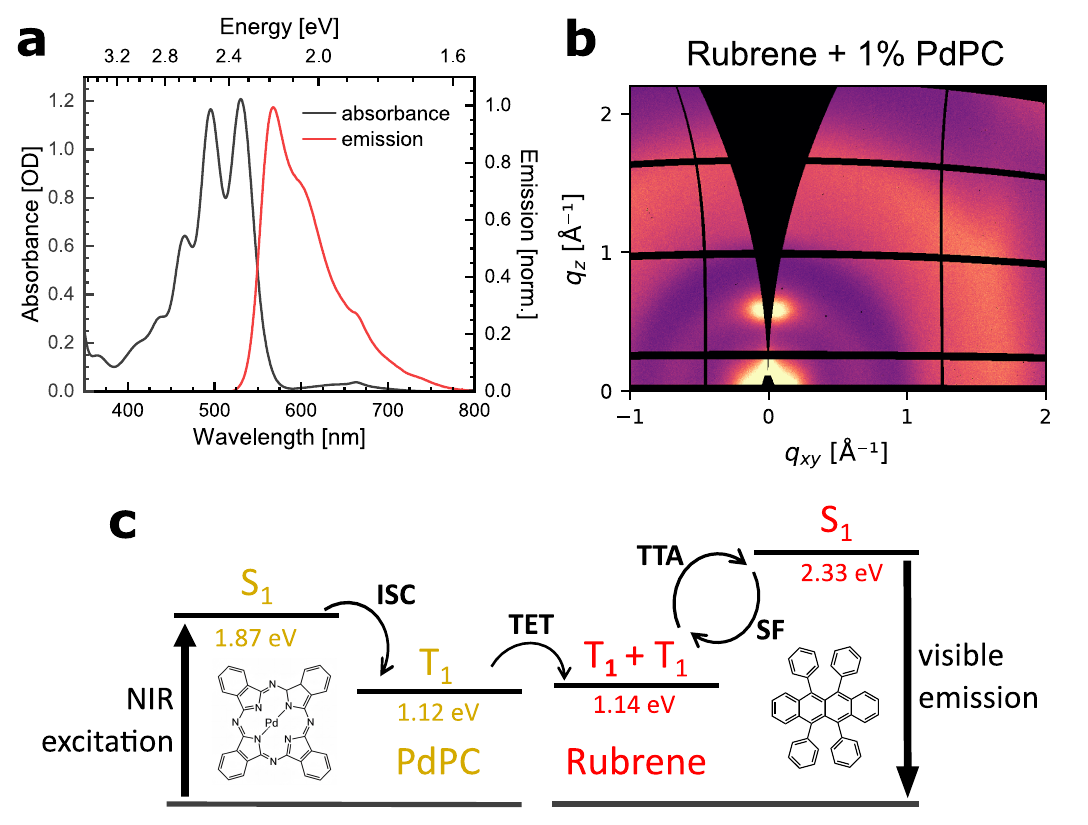}
       \caption{\textbf{Structural and photophysical characterisation of the rubrene:PdPC upconversion films.} \textbf{(a)} Absorption (grey) and photoluminescence (red, $\lambda_{\text{exc}} = 405$~nm) spectra for a 500~nm thick rubrene film doped with $\SI{1}{vol\%}$ PdPC prepared by thermal evaporation. \textbf{(b)} 2D grazing-incidence wide-angle X-ray scattering (GIWAXS) pattern of an identically fabricated rubrene:PdPC film. \textbf{(c)} Schematic illustrating the triplet--triplet annihilation upconversion (TTA-UC) mechanism. Low-energy excitation of the PdPC sensitizer ($\textrm{S}_1 = 1.87$~eV) is followed by efficient intersystem crossing (ISC) to the triplet state ($\textrm{T}_1 = 1.12$~eV). Subsequent triplet energy transfer (TET) to rubrene ($\textrm{T}_1 = 1.14$~eV) enables TTA to populate the rubrene singlet state ($\textrm{S}_1 = 2.33$~eV) which can radiatively decay (upconversion) or repopulate triplet states via singlet fission (SF).}
    \label{fig_c7_intro_pdpc_tta_schematic_abs_pl_giwaxs}
\end{figure}

We initially optimized the doping concentration (see SI Figure \ref{SI:fig_c7_SI_405 nm pl pdpc opti}), identifying that a low $\SI{1}{vol\%}$ doping concentration of PdPC yields the highest emission intensity as it limits non-radiative losses associated with aggregation;\cite{torrePhthalocyaninesOldDyes2007} but still maintains a relatively strong absorption for upconversion. To ensure both PdPC's absorption and rubrene's emission can be enhanced simultaneously, we have opted for a relatively thick rubrene-PdPC active layer ($\SI{500}{\nano\meter}$) to provide ample interaction with the electric-field maxima of two distinct resonant modes ($2\lambda$ and $\frac{5}{2}\lambda$, see SI Figure \ref{SI:PdPC_efields_SI} for electric field distributions). Furthermore, the thicker layer should provide a greater baseline PdPC absorption and thus upconversion signal. The absorption and emission spectra of this active layer are presented in Figure \ref{fig_c7_intro_pdpc_tta_schematic_abs_pl_giwaxs}a.

The doped films show a well-defined vibronic progression typically associated with disordered rubrene.\cite{irkhinAbsorptionPhotoluminescenceSpectroscopy2012} Here, the $S_{1}$ 0-0 absorption peak at $\SI{530}{\nano\meter}$ (\SI{2.34}{\electronvolt}) is approximately twice the energy of rubrene's triplet state (\SI{1.14}{\electronvolt}),\cite{herkstroeterTripletStateEnergies1981, yongEntangledTripletPair2017} making TTA energetically feasible. We also find a small absorption peak at $\SI{665}{\nano\meter}$ (\SI{1.87}{\electronvolt}) corresponding to PdPC's $\textrm{S}_1$ state that we aim to enhance.

To assess whether the addition of PdPC alters the structure of the rubrene films, we performed grazing-incidence wide-angle X-ray scattering (GIWAXS), as shown in Figure~\ref{fig_c7_intro_pdpc_tta_schematic_abs_pl_giwaxs}b. The scattering is dominated by broad, low-intensity features, consistent with a weakly ordered morphology. The rubrene-associated feature at $q = 0.59\text{~\AA}^{-1}$, corresponding to a characteristic spacing of $d = 2\pi/q = 10.6\text{~\AA}$, exhibits a pronounced preference for the out-of-plane $q_z$ direction, indicating preferential orientation despite the limited structural order. Compared with neat rubrene (SI Figure~\ref{SI:pure_rub_giwaxs_fig_SI}), the PdPC-containing film exhibits additional scattering at low $q$, consistent with the presence of larger-scale PdPC aggregates, together with an additional broad, diffuse contribution at higher $q$, indicative of short-range order. Despite these additional contributions, the rubrene-associated scattering remains very similar to that of the neat film. Scherrer analysis of the out-of-plane peak at $q_z = 0.59\text{~\AA}^{-1}$ yields comparable crystalline coherence lengths of $4.0 \pm 0.1\text{~nm}$ and $3.7 \pm 0.1\text{~nm}$ for the neat and PdPC-containing films, respectively. These observations suggest that, although PdPC contributes additional scattering, its inclusion does not substantially alter the local packing or crystalline coherence of rubrene.


The energetic separation between the $\textrm{S}_1$ states of rubrene and PdPC represents the maximum energy gain achievable through TTA-UC. To illustrate this, we have drawn an energy-level diagram (Figure \ref{fig_c7_intro_pdpc_tta_schematic_abs_pl_giwaxs}c) that summarizes the mechanism. Following NIR excitation of the PdPC $\textrm{S}_1$ state, the heavy palladium centre facilitates ISC to the $\textrm{T}_1$ state with a near-unity yield.\cite{radiunasUnderstandingLimitationsNIRtovisible2020a, hanFirstTransitionMetal2017, chePhotophysicalPropertiesPalladium2018} Given the close energetic alignment between the PdPC triplet (\SI{1.12}{\electronvolt})\cite{radiunasUnderstandingLimitationsNIRtovisible2020a} and the rubrene triplet manifold (\SI{1.14}{\electronvolt}),\cite{herkstroeterTripletStateEnergies1981, yongEntangledTripletPair2017} triplet energy transfer (TET) occurs. Subsequent bimolecular TTA within the rubrene host then populates its $\textrm{S}_1$ state, resulting in up-converted emission from rubrene at \SI{2.18}{\electronvolt}.

To achieve a resonant enhancement in both absorption and emission, whilst simultaneously achieving strong light-matter coupling, we use Ag mirrors as their reflectance remains consistently high across across these regions. Furthermore, the electric-field is tightly confined in planar Ag microcavities,\cite{maPhaseShiftPenetration2007} which is beneficial for both the resonant enhancements and strong coupling. Here, we opt for a relatively thin ($\SI{30}{\nano\meter}$) top mirror and thicker ($\SI{200}{\nano\meter}$) bottom mirror so we can probe the reflectance and ensure efficient out-coupling.\cite{linExaminingMicrocavityOrganic2005} Optically inert \ch{SiO2} spacers are then used to prevent non-radiative quenching typically associated with metal interfaces.\cite{nowyLightExtractionOptical2008, barnesFluorescenceInterfacesRole1998} A schematic of this device stack is presented in Figure \ref{fig_c7_pdpc_cavity_design}a:

\begin{figure}[H]
    \centering
    \includegraphics[width=1\linewidth]{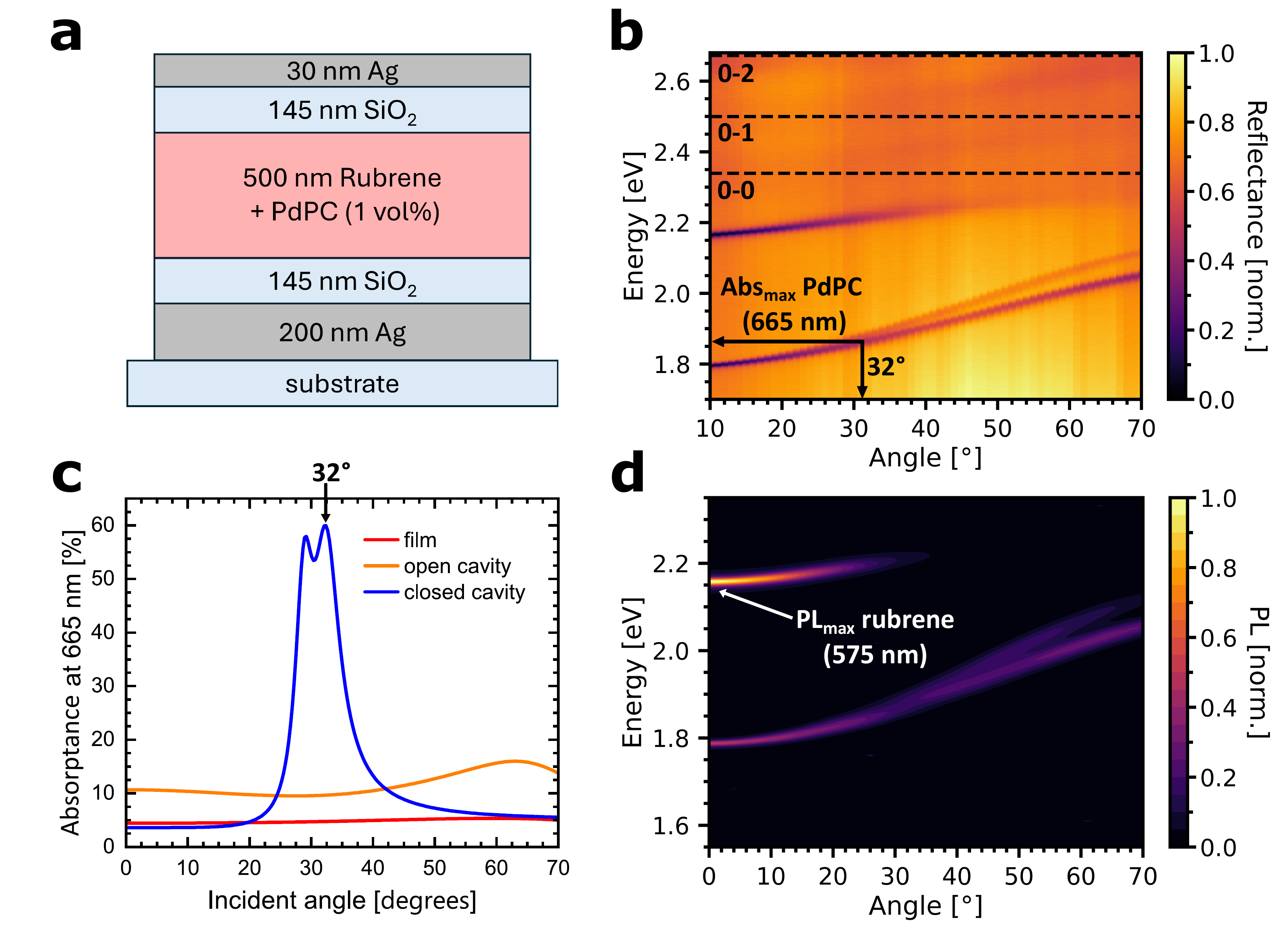}
       \caption{\textbf{Design and optical characterisation of the resonant closed microcavity structure.} \textbf{(a)} Schematic of the microcavity device stack featuring a 500~nm rubrene:PdPC active layer sandwiched between Ag mirrors and SiO$_2$ spacer layers. \textbf{(b)} Angle-resolved white-light reflectance colour map of the closed cavity. The dashed lines indicate the rubrene vibronic absorption transitions (0-0, 0-1, 0-2). The dispersive lower energy mode is tuned to enhance PdPC absorption at 32$^{\circ}$ and the higher energy mode is positioned to enhance rubrene emission at 0$^{\circ}$. \textbf{(c)} Angular absorptance at 665~nm, demonstrating a resonant absorption enhancement of the closed cavity at 32$^{\circ}$. \textbf{(d)} Angle-resolved photoluminescence (PL) map ($\lambda_{\text{exc}} = 405$~nm). The emission intensity is maximised when the lower polariton branch (LPB) is isoenergetic with rubrene's emission at 0$^{\circ}$.}
    \label{fig_c7_pdpc_cavity_design}
\end{figure}

For our unconfined controls, the closed microcavity structure will be compared to an open cavity (structure and angular reflectance characterisation shown in SI Figure \ref{SI:fig_c7_SIpdpc_open_cavity_design_SI}) in addition to a bare film. Both of these samples are encapsulated with \ch{SiO2} to avoid degradation of the active layer through oxygen exposure.\cite{kytkaRealtimeObservationOxidation2007} Importantly, the open cavity is essentially a back reflector, which is much simpler to fabricate than the closed cavity, but can still re-direct light broadly in the forward direction. Thus, for the microcavity to be beneficial to TTA-UC, it would need to out-perform the open cavity.

To verify whether the closed microcavity is functioning as intended, we have performed angular reflectance measurements to characterize the confined modes and examine the light--matter interaction (Figure \ref{fig_c7_pdpc_cavity_design}b). From inspection, two clear modes are supported by the cavity: a lower-energy mode positioned to enhance sensitizer absorption, and a higher-energy mode designed to maximize emission outcoupling.

The lower-energy cavity mode becomes isoenergetic with the PdPC absorption maximum (\SI{665}{\nano\meter}) at an incidence angle of \ang{32}, driving a 12-fold enhancement in absorptance over the bare control film. Transfer-matrix method (TMM) simulations of the microcavity absorptance at \SI{665}{\nano\meter} as a function of incidence angle illustrate this boost (Figure \ref{fig_c7_pdpc_cavity_design}c; see SI Figure \ref{SI:PdPC_abs_comparison_SI} for full spectra). Because sensitizer absorption peaks sharply at \ang{32}, we select this angle of incidence for excitation to maximize the overall upconversion signal.

As microcavity emission tends to be maximized in the forward direction, normal to the substrate surface,\cite{benistyImpactPlanarMicrocavity1998} we aim to collect upconverted light at this point. Accordingly, the higher-energy cavity mode is positioned to resonate with rubrene's 0-0 emission peak (\SI{575}{\nano\meter}) at a collection angle of \ang{0}. Angular emission measurements under \SI{405}{\nano\meter} excitation (Figure~\ref{fig_c7_pdpc_cavity_design}d) support this strategy, demonstrating maximum emission intensity at \ang{0} where the cavity mode overlaps spectrally with rubrene's emissive states. At higher angles, Figure~\ref{fig_c7_pdpc_cavity_design}b reveals this mode intersecting rubrene's absorption transitions (labeled 0-0, 0-1, and 0-2) and forming distinct anti-crossings, characteristic of the strong light--matter coupling. This behavior is further confirmed by simulated angular reflectance profiles (SI Figure~\ref{SI:PdPC_TMM_angR_strong_coupling_SI}). Under this coupling regime, the hybrid mode forms a lower polariton branch (LPB), through which upconverted emission is efficiently outcoupled via radiative pumping from uncoupled rubrene states.\cite{grantEfficientRadiativePumping2016}

The predicted enhancements in absorption and emission stem from the increased local electric field intensity and local density of optical states (LDOS) provided by photonic confinement. This combination accelerates transition rates within the active layer as described by Fermi’s Golden Rule.\cite{diracPrinciplesQuantumMechanics1981, fermiNuclearPhysicsCourse1950, visserWhoseGoldenRule2009} Indeed, simulated electric field distributions (SI Figure \ref{SI:PdPC_efields_SI}) confirm that the field is strongly concentrated within the active layer at both the absorption wavelengths of PdPC and the emission wavelengths of rubrene, producing a synergistic improvement in the total upconversion signal.

We now assess the upconversion measurements where PdPC is excited at \ang{32} with a $\SI{665}{\nano\meter}$ source (Fianium Supercontinuum laser) and emission is collected at a viewing angle of \ang{0}. The TTA-UC emission spectra of the film, the open cavity, and the closed cavity are presented in Figure \ref{fig_c7_PdPC_UCPL_spectra}:

\begin{figure}[H]
    \centering
    \includegraphics[width=0.7\linewidth]{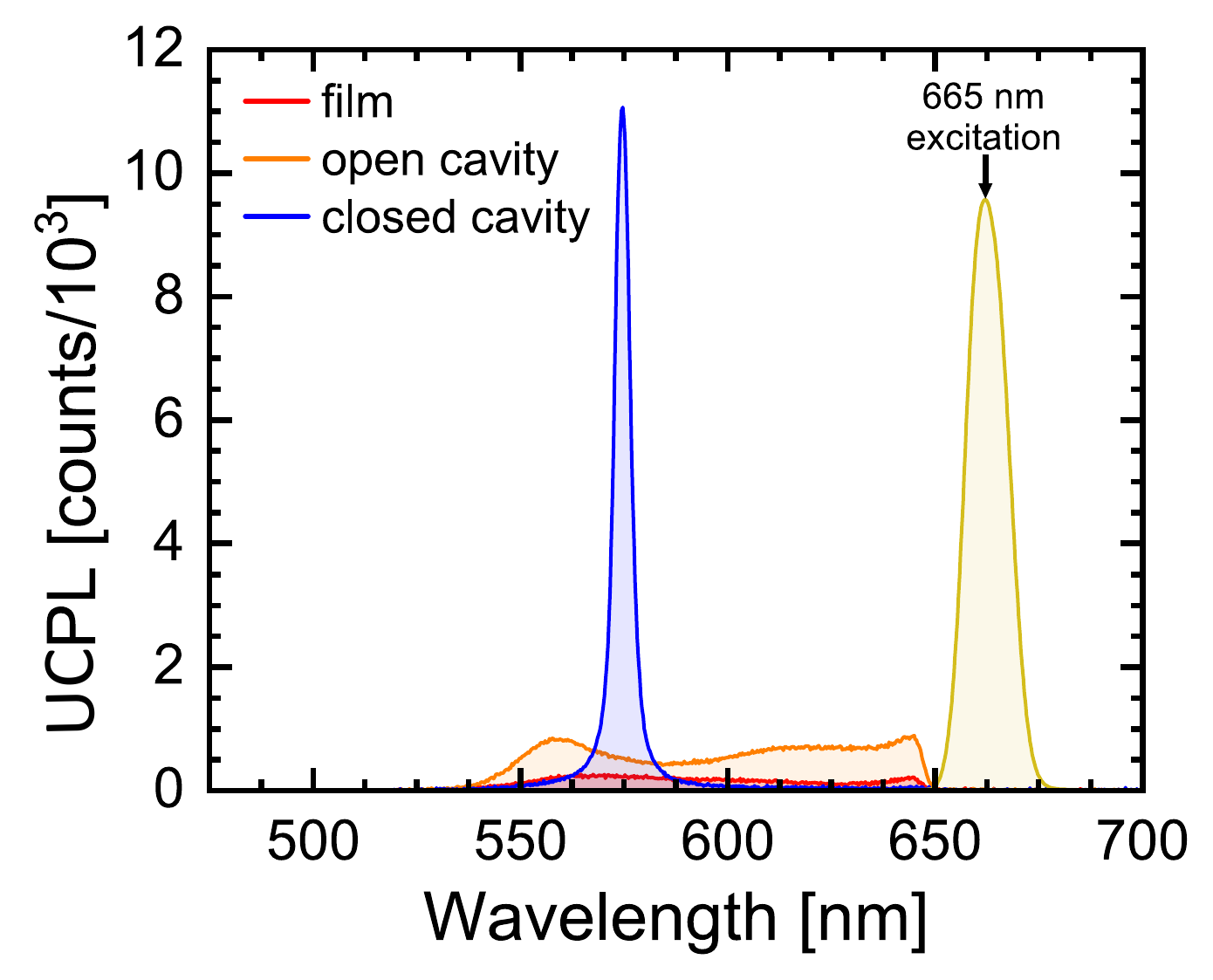}
       \caption{\textbf{Upconversion photoluminescence (UCPL) spectra of the rubrene:PdPC system in different architectures.} UCPL spectra for the control film (red), open cavity (orange), and closed cavity (blue) under 665~nm excitation ($P_{\text{exc}} = 7.7$~mW). Measurements were performed in a reflection geometry with emission collected at 0$^{\circ}$. The yellow shaded region represents the spectrum of the excitation source which has been scaled down in the plot to be comparable to the UCPL spectra.}
    \label{fig_c7_PdPC_UCPL_spectra}
\end{figure}

Examining the up-converted spectra, the closed cavity exhibits the highest peak intensity with a 22-fold relative enhancement over the film, suggesting that the microcavity design is working as intended. However, the film and open cavity spectra are notably broader, which must be taken into account. Due to the spectral proximity of the excitation source and the up-converted signal, the emission at longer wavelengths is somewhat distorted by minor laser leakage through the short-pass filter ($\lambda < \SI{650}{\nano\meter}$). To ensure an accurate comparison, we therefore evaluate the integrated emission intensity within the \SIrange{525}{625}{\nano\meter} range. At the highest power measured, we find the open and closed cavities enhance the upconversion signal by 3.3-fold and 4.5-fold over the bare film, respectively. The intermediate enhancement of the open cavity likely stems from back-reflection of the excitation source (enhancing PdPC absorption) and redirection of emission toward the detector. Despite the spectral narrowing inherent to the closed cavity, its substantial peak enhancement is sufficient to drive a superior overall signal. 

Considering that TTA is fundamentally a bimolecular process, further insight into the mechanism can be achieved by plotting the integrated emission signal against the incident excitation power ($P_{\text{exc}}$) on a double-logarithmic scale, which we present in Figure \ref{fig_c7_PdPC_UCPL_power}:

\begin{figure}[H]
    \centering
    \includegraphics[width=0.7\linewidth]{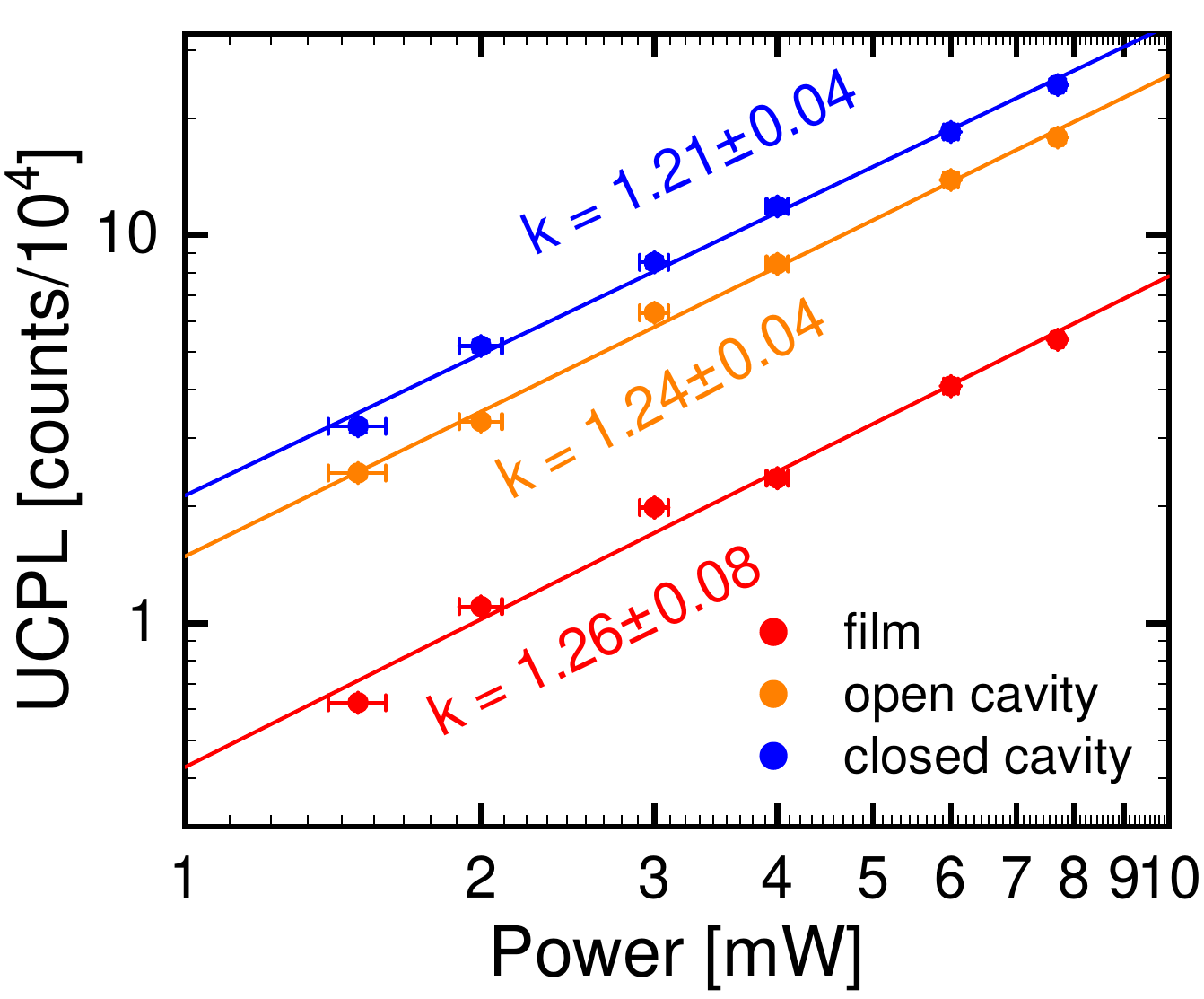}
       \caption{\textbf{Upconversion photoluminescence (UCPL) excitation power dependence.} Double-logarithmic plot of the integrated UCPL intensity ($\lambda_{\text{UCPL}} = 525$--625~nm) as a function of incident excitation power ($\lambda_{\text{exc}} = 665$~nm). The accessible fluence range (50--250~mW~cm$^{-2}$) is limited by the relatively large excitation spot size (\SI{2}{mm} diameter) used for these measurements. The error of the excitation power is derived from the laser-power fluctuation, and the UCPL error bars are based on the standard error of three separate, but identically fabricated, repeat samples measured at the highest power. The gradients of the straight line fits, $k$, represents the power-law index and indicates that triplets decay predominantly via TTA in these systems.}
    \label{fig_c7_PdPC_UCPL_power}
\end{figure}

Here we fit the data for each sample with a linear regression, where the gradient and intercept (at $P_{\text{exc}} = \SI{1.0}{mW}$) provide information about the excitation threshold and the efficiency of the systems. Following the steady-state rate equations for a two-photon process within a fixed excitation volume, the upconversion intensity ($I_{\text{UC}}$) scales with the absorbed power ($P_{\text{abs}}$) as:\cite{murakamiKineticsPhotonUpconversion2021}
\begin{equation}
I_{\text{UC}} = \eta (P_{\text{abs}})^k = \eta (A \cdot P_{\text{exc}})^k
\label{eq:tta_power_law}
\end{equation}
where $A$ is the absorptance of the sensitizer at the excitation wavelength, and $\eta$ is a proportionality constant representing the upconversion efficiency of the system specific to our measurement setup. The power-law index, $k$, identifies the dominant triplet decay regime: $k \to 2$ indicates the ``weak" regime where triplets decay primarily via mono-molecular non-radiative pathways such as triplet-back transfer\cite{linStrategiesHighPerformanceSolidState2020a} or quenching at defect sites.\cite{irkhinDefectDensityDependent2016} In contrast, $k \to 1$ denotes the ``strong" annihilation limit where TTA becomes the primary fate of the triplet population.

Taking the logarithm of Equation~\ref{eq:tta_power_law}:
\begin{equation}
\log_{10}(I_{\text{UC}}) = k \log_{10}(P_{\text{exc}}) + \log_{10}(\eta) + k \log_{10}(A)
\label{eq:tta_power_law_logged}
\end{equation}
reveals that the gradients of the linear fits in Figure~\ref{fig_c7_PdPC_UCPL_power}b represent the power-law index, $k$, for each sample. Across an average excitation power range of 1.5--7.7~mW (50--250~mW~cm$^{-2}$), we extract $k = 1.26 \pm 0.08$ for the bare film, $1.24 \pm 0.04$ for the open cavity, and $1.21 \pm 0.04$ for the closed cavity. These values deviate from the weak-annihilation limit ($k = 2$), indicating that TTA is already a dominant triplet decay channel across this excitation regime. 

The subtle reduction in $k$ from the bare film to the closed cavity, accompanied by a vertical offset in total UCPL intensity, reflects an increased steady-state triplet concentration driven by cavity-enhanced absorption. This close proximity to the strong-annihilation limit at modest average fluences (16--250~mW~cm$^{-2}$) compares favorably to literature values for similar rubrene:PdPC systems, where threshold intensities ($I_{\text{th}}$) of $1400\text{~mW~cm}^{-2}$ have been reported.\cite{radiunasEfficientNIRtovisPhoton2022a} However, our $40\text{~MHz}$ pulsed excitation source may contribute to lower effective threshold fluences despite operating in a quasi-continuous-wave regime ($25\text{~ns}$ pulse period $\ll\tau_T$).\cite{yeKineticsEfficiencyTripletTriplet2021}

Despite these relatively efficient internal kinetics, the absolute upconversion signal remains low, preventing a measurement of the upconversion yield with an integrating sphere. This indicates that the bottlenecks occur predominantly post-annihilation, likely due to the singlet fission (SF) pathway inherent to rubrene and detrimental self-absorption effects associated with the thick active layer. Consequently, we evaluate the upconversion efficiencies ($\eta$) specific to our experimental setup, which we can infer from the intercepts of the logarithmic plots now that we have calculated the power-law index for each sample and have knowledge of their absorptances (see Figure \ref{fig_c7_pdpc_cavity_design}c). The upconversion efficiency of each sample can then be divided by that of the film, giving us a gauge of how each sample is performing relative to the bare film. These values are summarized in Table \ref{tab_UC:UCPL_enhancement_external_efficiency}:

\begin{table}[H]
\centering
\caption{\textbf{Summary of upconversion performance parameters for rubrene:PdPC films and microcavity architectures.} Absorptance values ($A$) at \SI{665}{\nano\meter} were determined from transfer-matrix method simulations. Relative upconversion intensities ($I_{\text{UC}}$) were calculated from integrated emission spectra and normalized to the control film, with uncertainties representing the standard error across three identically fabricated repeat samples propagated in quadrature. Power-law indices ($k$) and intercepts, along with their associated uncertainties, were extracted directly from linear regression fits. Relative upconversion efficiencies ($\eta_{\text{rel}}$) were determined by evaluating $\eta$ for each sample (Equation~\ref{eq:tta_power_law_logged}, $\text{intercept} = \log_{10}(\eta) + k \log_{10}(A)$) and normalizing to the control film; reported errors reflect the full propagation in quadrature of individual sample and film uncertainties.}
\label{tab_UC:UCPL_enhancement_external_efficiency}
\renewcommand{\arraystretch}{1.3} 
\begin{tabular}{l c c c}
\toprule
\textbf{Parameter} & \textbf{Film} & \textbf{Open cavity} & \textbf{Closed cavity} \\
\midrule
Absorptance at \SI{665}{\nano\meter}, $A$ [\%] & 5    & 10   & 60   \\
Relative upconversion intensity, $I_{\text{UC}}$ & $1.0$ & $3.3\pm0.1$ & $4.5\pm0.2$ \\
Power-law index, $k$ & $1.26\pm0.08$ & $1.24\pm0.04$ & $1.21\pm0.04$ \\
Intercept [$\log_{10}(\text{counts})$]               & $3.63\pm0.05$ & $4.17\pm0.02$ & $4.32\pm0.03$ \\
Relative upconversion efficiency, $\eta_{rel}$     & $1.0$ & $1.5\pm0.3$ & $0.26\pm0.07$ \\
\bottomrule
\end{tabular}
\end{table}

In Table~\ref{tab_UC:UCPL_enhancement_external_efficiency}, we find that the closed cavity exhibits a reduced relative upconversion efficiency ($\eta_{\text{rel}} = 0.26 \pm 0.07$). This is a consequence of poor outcoupling stemming from the substantial self-absorption imposed by optical confinement, where the simultaneous enhancement of absorption and emission at the LPB position inherently increases photon reabsorption probability. In contrast, the open cavity demonstrates an enhanced efficiency ($\eta_{\text{rel}} = 1.5 \pm 0.3$), as its improved outcoupling toward the detector is not offset by severe reabsorption. To mitigate these self-absorption losses in future cavity architectures, the active-layer thickness and sensitizer doping concentration can be systematically optimized. Nevertheless, the absolute upconversion intensity remains highest in the closed cavity due to the considerable boost in absorptance ($A = 60\%$) and peak emission driven by the dual-resonator design.

In summary, while microcavity enhancement trades broad-band response for an intense, narrow-band resonant boost within a specific excitation and detection geometry, this photonic approach offers distinct advantages depending on the target application. The resulting narrow-band enhancement is particularly well suited for specialized technologies such as sub-diffraction-limited 3D printing,\cite{limbergTripletTripletAnnihilation2022} high-sensitivity anti-counterfeiting tags,\cite{narayananAlleviatingParasiticBack2025} or wavelength-selective near-infrared sensors.\cite{jiSemiconductorPlasmonEnhanced2019} Furthermore, by optimizing active-layer thicknesses to reduce reabsorption, microcavity-based TTA-UC provides a viable pathway for integration beneath wide-bandgap photovoltaics in tandem configurations to surpass fundamental power conversion efficiency limits.\cite{tayebjeeShockleyqueisserMolecularApproaches2015}

\begin{acknowledgement}

R.H.G. acknowledges the EPSRC and the University of Sheffield for funding via Impact Acceleration Account (EP/X525790/1). R.C.K. acknowledges the EPSRC for funding via grant EP/Y031962/1. The authors thank Diamond Light Source for the award of a block allocation group beamtime at the I07 beamline using the MINERVA chamber (experimental number: SI36456-2) and beamline scientist Jonathan Rawle for assistance with the experimental setup. The authors also thank Rahul Jayaprakash for providing the transfer matrix modeling code used to design the microcavity architectures and simulate their properties.





\end{acknowledgement}

\section*{Data Availability}
Data will be available on the University of Sheffield’s repository, ORDA.

\begin{suppinfo}


The Supporting Information is available free of charge on the ACS Publications website at DOI: [Insert DOI].

Experimental methods; sensitizer optimization; microcavity TMM simulations; and GIWAXS measurements (Insert PDF download link).

\end{suppinfo}



\clearpage

\renewcommand{\thepage}{S\arabic{page}}
\setcounter{page}{1}
\renewcommand{\thesection}{} 
\setcounter{section}{1}
\renewcommand{\thesubsection}{S\arabic{subsection}}
\renewcommand{\thetable}{S\arabic{table}}
\setcounter{table}{1}
\renewcommand{\thefigure}{S\arabic{figure}}
\setcounter{figure}{1}
\renewcommand{\theequation}{S\arabic{equation}}
\setcounter{equation}{1}


\section*{Supplementary information (SI)}

\subsection{Experimental Methods}\label{SI:SI_sec_Experimental_methods}

\subsubsection{Rubrene-PdPC Thin Film Fabrication.} All materials were deposited onto clean quartz-coated glass (S151, Ossila) substrates following sequential sonication for \SI{10}{min} each in a Hellmanex~III/deionized (DI) water solution, pure DI water, and isopropyl alcohol. The substrates were then dried under a continuous nitrogen stream and subjected to a \SI{10}{min} UV-ozone treatment. 500 nm rubrene thin films ($>99\%$, Sigma Aldrich) doped with 1 vol\% PdPC ($>95\%$, PorphyChem) were subsequently deposited via thermal evaporation (Angstrom Engineering) under high vacuum ($\leq\SI{2e-6}{mbar}$). The PdPC doping concentration was controlled by the relative deposition rates of rubrene (\SI{0.2}{nm.s^{-1}}) and PdPC (\SI{0.002}{nm.s^{-1}}), which were monitored simultaneously using two quartz crystal microbalances, calibrated by tooling factors verified by a Bruker DektakXT stylus profilometer. The sublimation temperatures of rubrene and PdPC were approximately \SI{190}{\degree C} and \SI{400}{\degree C}, respectively. To prevent photo-oxidation, all bare films and open cavity controls were immediately encapsulated with a \SI{150}{nm} layer of $\text{SiO}_2$ deposited via electron-beam evaporation.

\subsubsection{Microcavity Design and Fabrication.} Planar microcavities were designed using a custom transfer matrix method (TMM) model to simulate multi-interfacial reflections and optimize layer thicknesses. The complex refractive indices ($\tilde{N} = n + ik$) of the organic active layers were determined by mapping the experimental extinction coefficient ($k$) with Lorentzian oscillators, with the dispersive real component ($n$) calculated via a Kramers-Kronig transform. Optical constants for the silver mirror layers were verified via spectroscopic ellipsometry (J.A. Woollam M-2000) at an incidence angle of \SI{75}{\degree}. Silver (Ag) mirrors were deposited via thermal evaporation at a base pressure of \SI{2e-6}{mbar} using a vacuum deposition system (Angstrom Engineering). Here, Ag pellets were added to a resistive boat, which was heated with an electrical current, causing the silver to evaporate onto the samples suspended above. The thickness and deposition rates were monitored via quartz crystal microbalance and calibrated using a Bruker DektakXT stylus profilometer. Using this method, bottom Ag mirrors were deposited at a thickness of \SI{200}{nm} to act as a broadband reflector, while top mirrors were deposited at \SI{35}{nm} to allow for sufficient out-coupling of light.

\subsubsection{Grazing-Incidence Wide-Angle X-ray Scattering (GIWAXS).} Grazing-incidence wide-angle X-ray scattering (GIWAXS) measurements were performed at the I07 surface diffraction beamline at Diamond Light Source using an incident X-ray energy of $20\text{~keV}$ ($\lambda = 0.62\text{~\AA}$). Samples were deposited onto Si(100) substrates to reduce amorphous background scattering and housed in the Microstructure and Interface Evolution in Realtime under VAcuum (MINERVA) chamber, maintained at a base pressure of approximately $1 \times 10^{-7}\text{~mbar}$.\cite{nicklinMINERVAFacilityStudy2017} The incidence angle was fixed at \ang{0.075}. Scattered X-rays were recorded using a Pilatus 2M detector positioned approximately $0.41\text{~m}$ from the sample. The sample-to-detector distance was calibrated using a silver behenate standard in grazing-incidence geometry. The raw two-dimensional scattering images were geometrically corrected, masked to exclude regions shadowed by the MINERVA chamber window ports, and mapped into reciprocal-space coordinates using the \texttt{pyFAI} and \texttt{pygix} Python packages.\cite{ashiotisFastAzimuthalIntegration2015} Rubrene crystalline coherence lengths (CCLs) were estimated by fitting a Lorentzian function to the out-of-plane peak at $q_z = 0.59\text{~\AA}^{-1}$ and applying the Scherrer equation,
\begin{equation}
\text{CCL} = \frac{2\pi K}{\Delta q}
\end{equation}
where $K$ is a dimensionless shape factor (taken as $0.9$) and $\Delta q$ is the full width at half maximum of the fitted peak in reciprocal-space units.


\subsubsection{Steady-State Optical Characterization.} Steady-state UV-vis absorption spectra of bare thin films were acquired using an Agilent Cary 60 spectrophotometer. Substrate baseline corrections and dark-current subtractions were performed to account for surface reflections and electronic noise, respectively.

Angle-resolved optical characterization of microcavities was conducted using a custom-built goniometer setup. For angle-resolved specular reflectance maps, collimated, unpolarized light from a deuterium-tungsten lamp (Ocean Optics DH-2000-BAL) was focused onto the sample at the center of the goniometer. Reflected light was collected from \ang{10} to \ang{70} in \ang{1} steps via a detection arm fiber-coupled to an Andor Shamrock SR-303i-A triple-grating CCD spectrometer. Reflectance spectra were calibrated against a planar silver reference mirror to establish absolute reflectance prior to normalization.

Angle-resolved photoluminescence (PL) spectra were recorded on the same goniometer. A continuous-wave \SI{405}{nm} diode laser (Thorlabs) focused via an out-of-plane excitation arm (incident angle \ang{40}) served as the excitation source. PL emission was collected over angles ranging from normal incidence at \ang{0} to \ang{70}. A long-pass filter ($\lambda > \SI{450}{nm}$) was positioned immediately prior to the spectrometer entrance to eliminate scattered excitation light.

Photon upconversion measurements of PdPC-sensitized films were performed using the same goniometer setup, substituting the diode laser with a supercontinuum laser (Fianium Femtopower 1060 with a SuperK VARIA tunable filter, \SIrange{400}{840}{nm}). To approximate continuous-wave excitation conditions while maximizing power throughput, the laser was operated at a repetition rate of \SI{40}{MHz}, with incident power tuned via a variable neutral density filter wheel. To isolate the upconverted anti-Stokes emission, a long-pass filter was inserted into the excitation beam prior to the sample to selectively excite the PdPC sensitizer without directly pumping the emitter. A short-pass filter was then placed in the collection arm prior to the spectrometer to reject scattered excitation light while selectively transmitting the higher-energy upconverted emission.

\subsection{PdPC emission quenching}\label{SI:C7_sec_SI_405 nm pl pdpc opti}

\begin{figure}[H]
    \centering
    \includegraphics[width=1\linewidth]{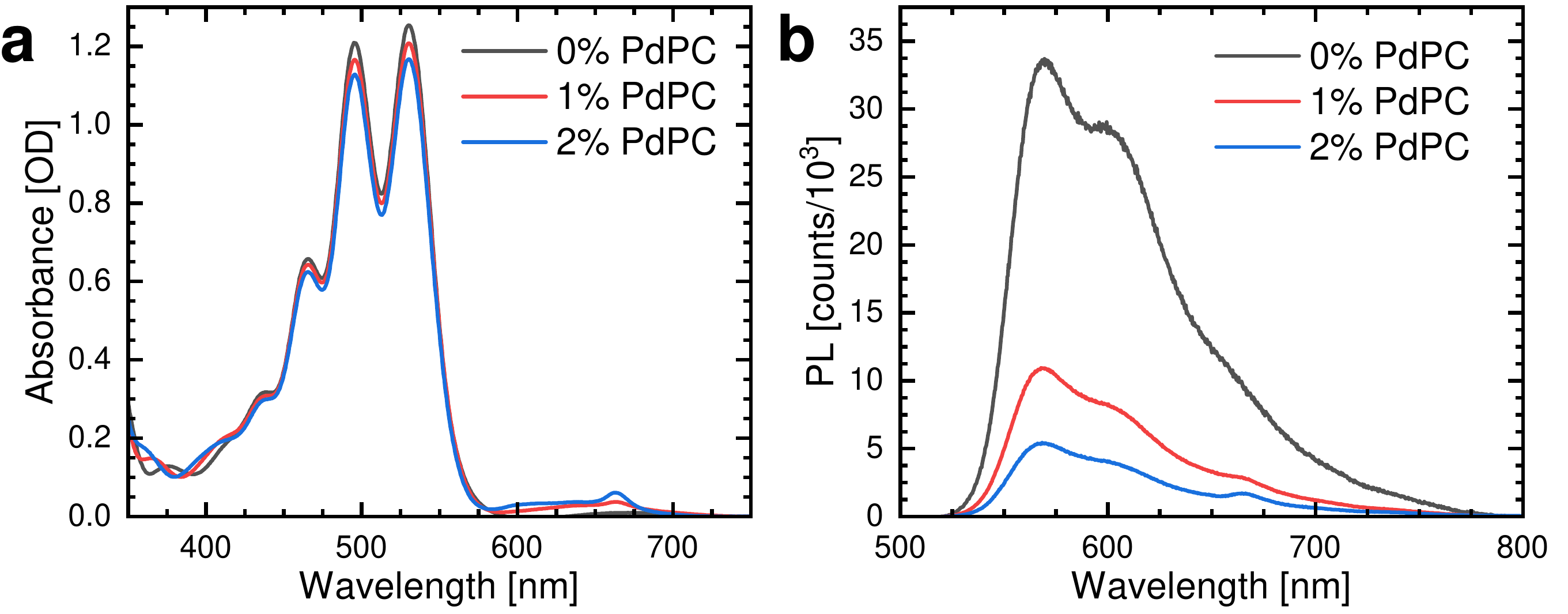}
       \caption{\textbf{Effect of PdPC doping concentration on the steady-state photophysics of rubrene films.} \textbf{(a)} Absorption spectra of rubrene films with varying PdPC doping concentrations. The absorption intensity of PdPC at 665~nm increases linearly with higher doping concentrations. \textbf{(b)} Photoluminescence (PL) spectra ($\lambda_{\text{exc}} = 405$~nm). Increasing the PdPC concentration leads to suppression of the rubrene emission, while a relative increase in PdPC emission is observed at 665~nm. This emission quenching at higher sensitiser concentrations suggests increased non-radiative losses, potentially arising from absorption of PdPC or singlet/triplet back-transfer from the rubrene host.}
    \label{SI:fig_c7_SI_405 nm pl pdpc opti}
\end{figure}

\subsection{Microcavity electric field distributions}\label{SI:C7_sec_SI_efields}

\begin{figure}[H]
    \centering
    \includegraphics[width=1\linewidth]{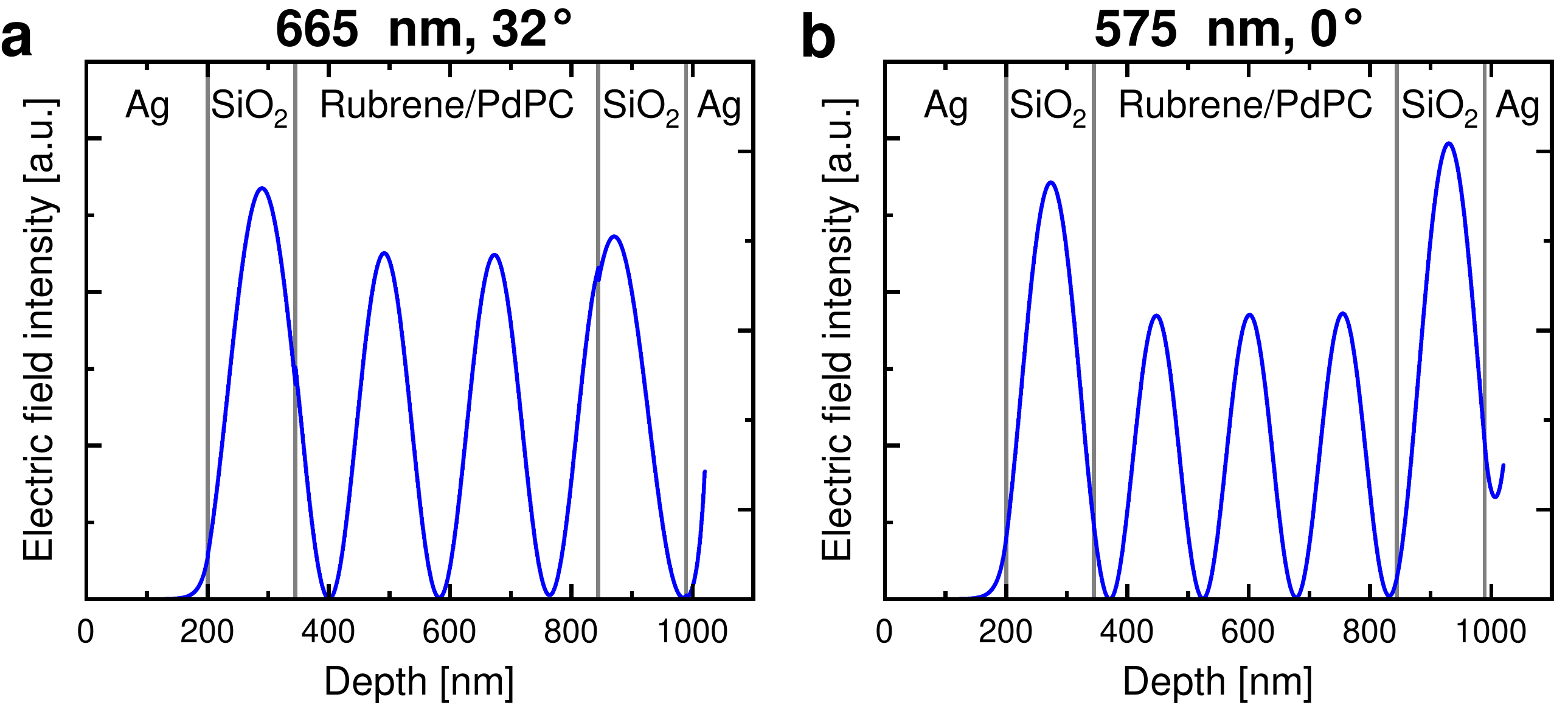}
       \caption{\textbf{TMM simulations of the electric field distribution within the closed microcavity.} \textbf{(a)} Electric field intensity profile at the excitation resonance (\SI{665}{nm} at 32$^{\circ}$). The profile exhibits four distinct peaks ($4\lambda/2$) confined within the cavity, with the rubrene:PdPC active layer positioned to overlap with the majority of these high-intensity regions to maximise absorption. \textbf{(b)} Electric field intensity profile at the emission resonance (\SI{575}{nm} at 0$^{\circ}$). This shows a fifth-order mode ($5\lambda/2$), where the electric field intensity is again high within the active layer region. This spatial overlap between the cavity modes and the active material is designed to facilitate the simultaneous resonant enhancement of both the sensitiser excitation and the up-converted rubrene emission.}
    \label{SI:PdPC_efields_SI}
\end{figure}

\subsection{Thermally evaporated rubrene X-ray scattering}\label{SI:pure_rub_giwaxs_sec_SI}

\begin{figure}[H]
    \centering
    \includegraphics[width=1\linewidth]{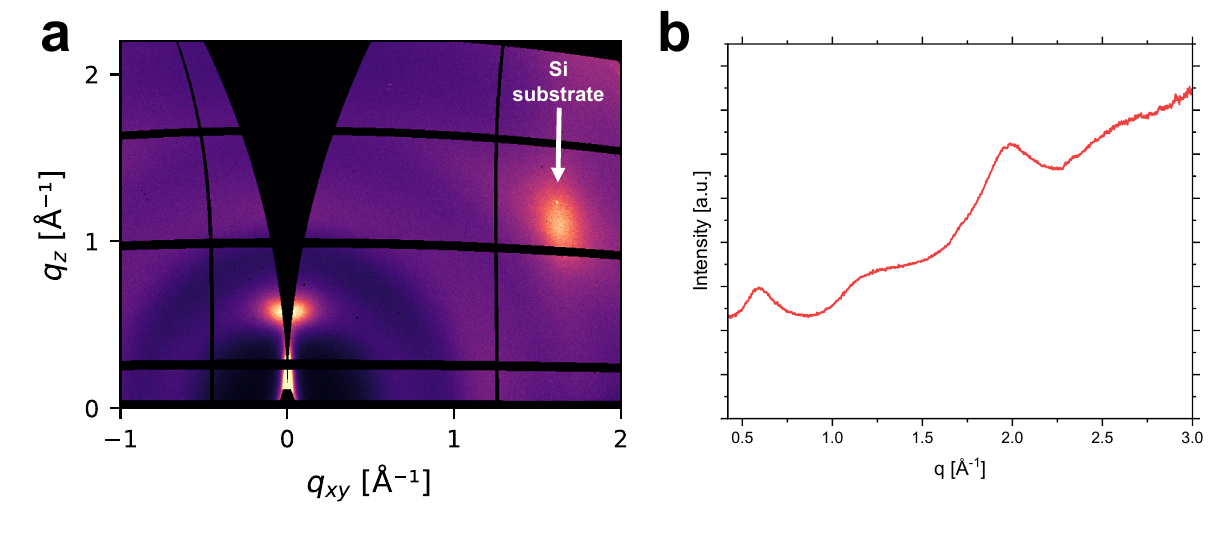}
       \caption{\textbf{Grazing-incidence wide-angle X-ray scattering (GIWAXS) characterization of a thermally evaporated neat rubrene film.} (a)~GIWAXS pattern showing a weakly ordered structure, with a pronounced out-of-plane preference for the feature at $q = 0.59\text{~\AA}^{-1}$. The broad spot at $q \approx 1.8\text{~\AA}^{-1}$ originates from the Si substrate. (b)~Scattering intensity profile as a function of $q$, obtained by azimuthal integration over \ang{0}--\ang{90} on the detector. The rubrene-associated scattering is very similar to that of the PdPC-containing film shown in the main text, despite the additional scattering contributions observed upon PdPC addition. Scherrer analysis of the out-of-plane rubrene peak yields comparable crystalline coherence lengths of $4.0\text{~nm}$ for neat rubrene and $3.7\text{~nm}$ for rubrene with PdPC, suggesting that PdPC inclusion does not substantially alter the local packing or crystalline coherence of rubrene.}
    \label{SI:pure_rub_giwaxs_fig_SI}
\end{figure}

\subsection{PdPC open cavity design and characterisation}\label{SI:C7_sec_SI_open_cavity}

\begin{figure}[H]
    \centering
    \includegraphics[width=1\linewidth]{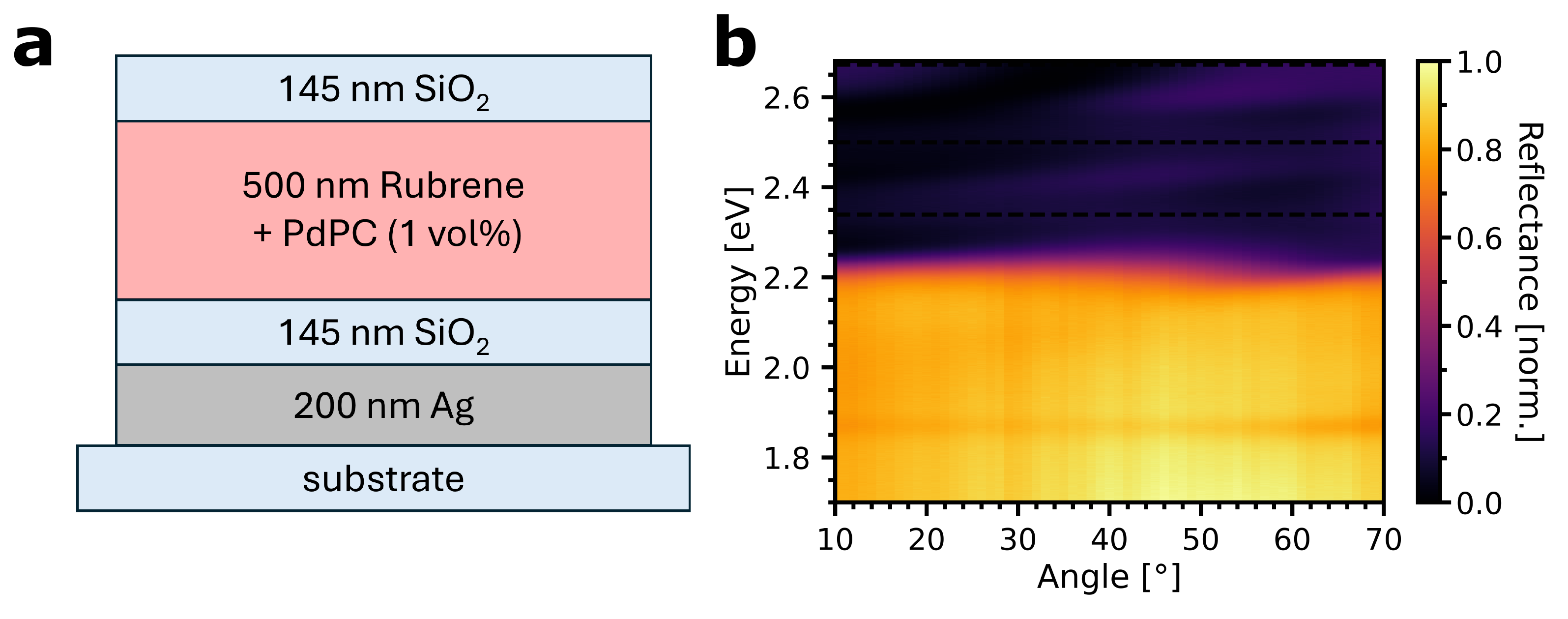}
       \caption{\textbf{Structure and characterisation of the open cavity.} \textbf{(a)} Schematic of the open cavity device stack, consisting of a $200\text{ nm}$ Ag back-reflector, a $145\text{ nm}$ SiO$_2$ spacer, and the $500\text{ nm}$ rubrene:PdPC ($1\text{ vol\%}$) active layer capped with a final  $145\text{ nm}$ SiO$_2$ layer. \textbf{(b)} Angle-resolved white-light reflectance map of the open cavity. Horizontal dashed lines indicate the rubrene vibronic absorption transitions, which correspond to regions of reduced reflectance. A further reflectance dip is observed at \SI{1.84}{eV}, corresponding to the absorption of the PdPC sensitiser.}
    \label{SI:fig_c7_SIpdpc_open_cavity_design_SI}
\end{figure}

\subsection{PdPC closed cavity supporting information}\label{SI:C7_sec_SI_cavity_design}

\begin{figure}[H]
    \centering
    \includegraphics[width=0.8\linewidth]{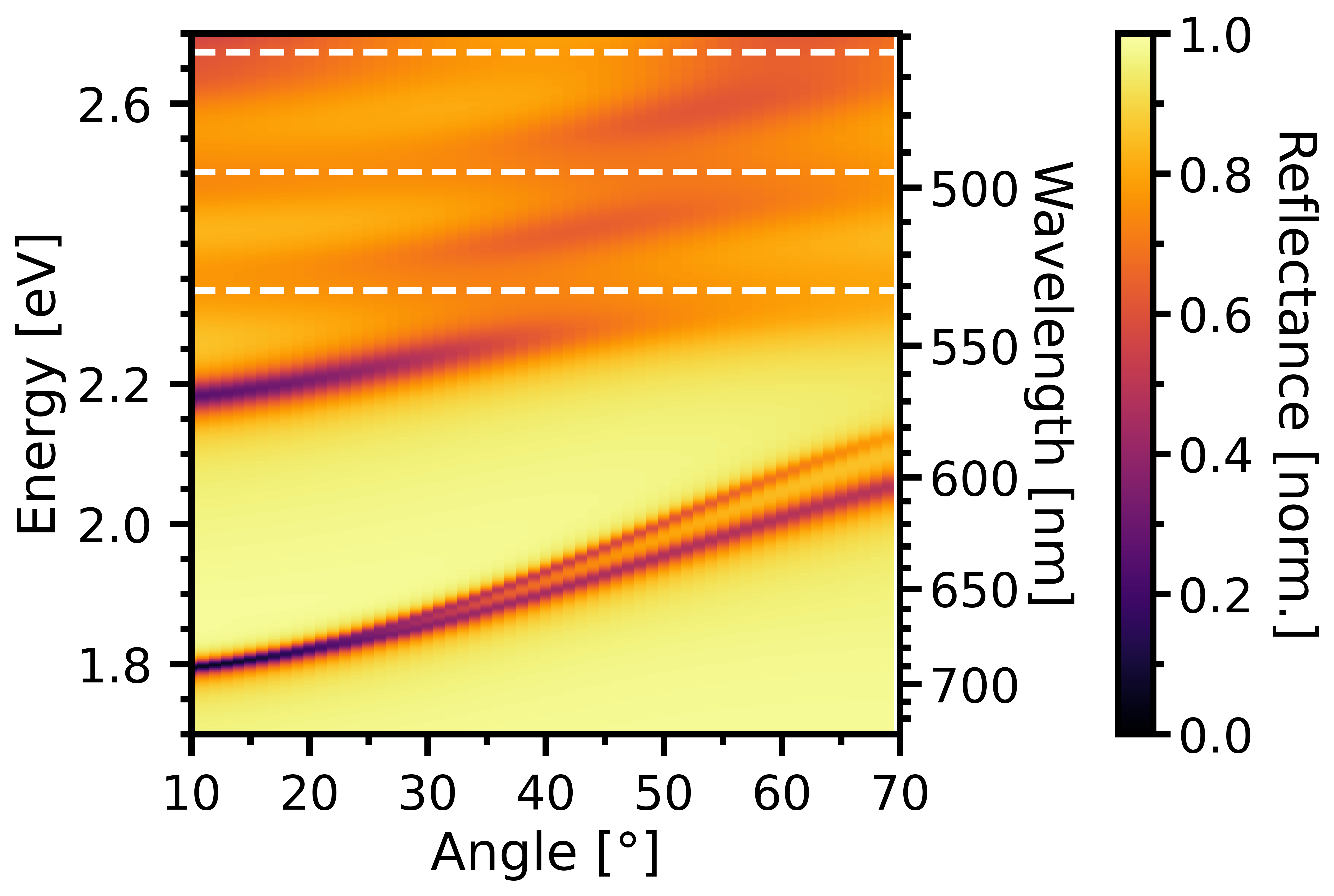}
       \caption{\textbf{Transfer matrix method (TMM) simulation of the closed cavity angular reflectance.} The simulation reveals anti-crossing between the cavity mode and the rubrene absorption transitions (dashed white lines), indicating the formation of hybrid light-matter states (polaritons).}
    \label{SI:PdPC_TMM_angR_strong_coupling_SI}
\end{figure}

\begin{figure}[H]
    \centering
    \includegraphics[width=0.75\linewidth]{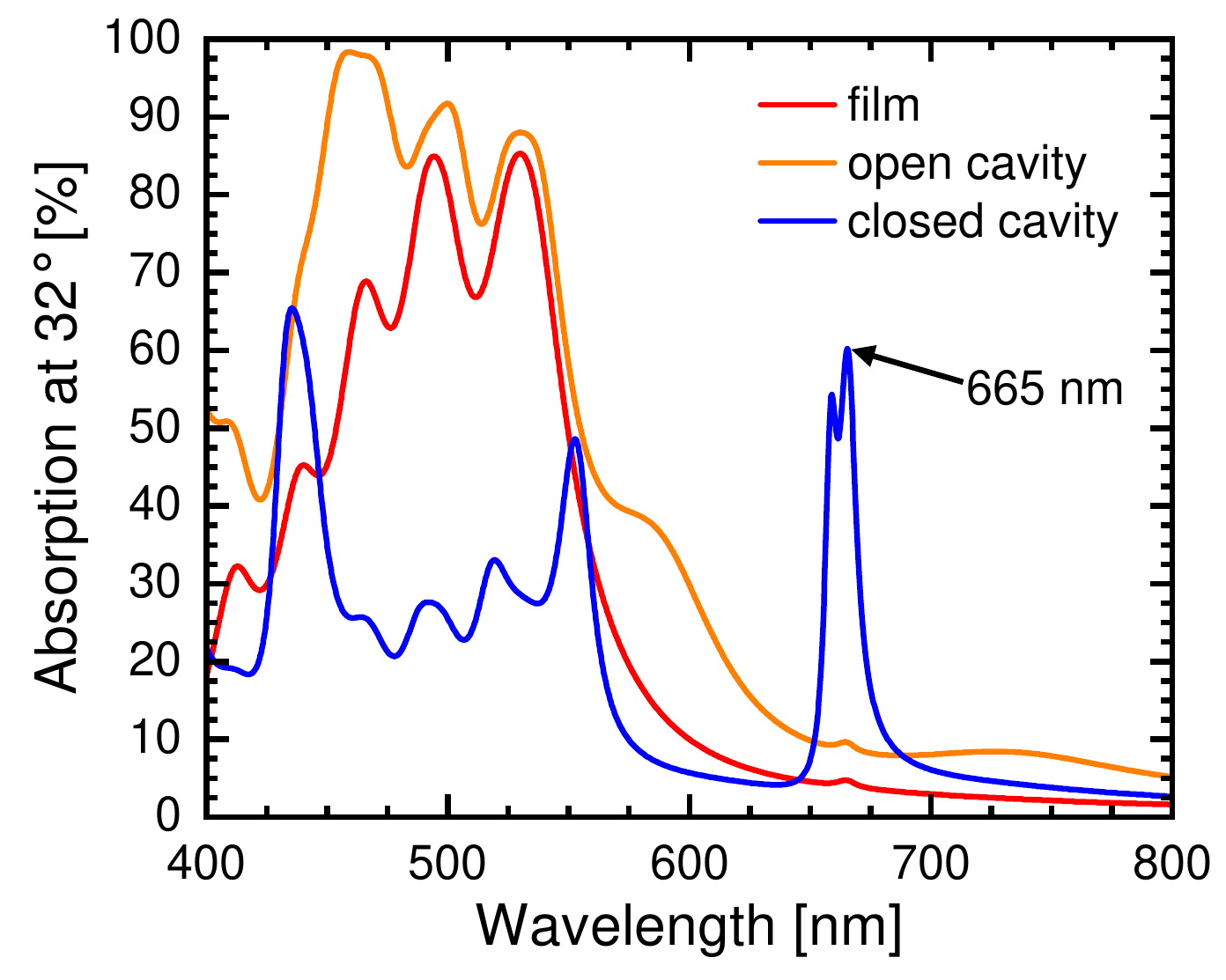}
       \caption{\textbf{Transfer matrix method (TMM) simulation of angular absorptance at 32$^{\circ}$.} Simulated absorptance spectra for the control film (red), open cavity (orange), and closed cavity (blue) at the optimised incident angle of 32$^{\circ}$. At the excitation wavelength (\SI{665}{nm}), the PdPC absorption is resonantly enhanced in the closed cavity architecture, reaching an absorptance of approximately 60\%. This represents a significant increase compared to the open cavity (10\%) and the bare film (5\%).}
    \label{SI:PdPC_abs_comparison_SI}
\end{figure}

\bibliography{library}

@article{grantEfficientRadiativePumping2016,
  title = {Efficient {{Radiative Pumping}} of {{Polaritons}} in a {{Strongly Coupled Microcavity}} by a {{Fluorescent Molecular Dye}}},
  author = {Grant, Richard T. and Michetti, Paolo and Musser, Andrew J. and Gregoire, Pascal and Virgili, Tersilla and Vella, Eleonora and Cavazzini, Marco and Georgiou, Kyriacos and Galeotti, Francesco and Clark, Caspar and Clark, Jenny and Silva, Carlos and Lidzey, David G.},
  year = {2016},
  journal = {Advanced Optical Materials},
  volume = {4},
  number = {10},
  pages = {1615--1623},
  issn = {21951071},
  doi = {10.1002/adom.201600337}
}

@article{tayebjeeShockleyqueisserMolecularApproaches2015,
  title = {Beyond Shockley-Queisser: {{Molecular}} Approaches to High-Efficiency Photovoltaics},
  author = {Tayebjee, Murad J.Y. and McCamey, Dane R. and Schmidt, Timothy W.},
  year = {2015},
  journal = {Journal of Physical Chemistry Letters},
  volume = {6},
  number = {12},
  pages = {2367--2378},
  publisher = {American Chemical Society},
  issn = {19487185},
  doi = {10.1021/acs.jpclett.5b00716}
}

@article{ye2021,
  title = {Direct {{Transition}} from {{Triplet Excitons}} to {{Hybrid Light}}–{{Matter States}} via {{Triplet}}–{{Triplet Annihilation}}},
  author = {Ye, Chen and Mallick, Suman and Hertzog, Manuel and Kowalewski, Markus and Börjesson, Karl},
  year = {2021},
  journal = {Journal of the American Chemical Society},
  shortjournal = {J. Am. Chem. Soc.},
  volume = {143},
  number = {19},
  pages = {7501--7508},
  publisher = {American Chemical Society},
  issn = {0002-7863},
  doi = {10.1021/jacs.1c02306},
  url = {https://doi.org/10.1021/jacs.1c02306},
  urlaccess = {2026-01-29}
}

@article{herkstroeterTripletStateEnergies1981,
  title = {The Triplet State Energies of Rubrene and Diphenylisobenzofuran},
  author = {Herkstroeter, W. G. and Merkel, P. B.},
  year = {1981},
  journal = {Journal of Photochemistry},
  shortjournal = {Journal of Photochemistry},
  volume = {16},
  number = {3},
  pages = {331--342},
  issn = {0047-2670},
  doi = {10.1016/0047-2670(81)80043-3},
  url = {https://www.sciencedirect.com/science/article/pii/0047267081800433},
  urlaccess = {2025-10-06}
}

@article{singh-rachfordPdIIPhthalocyanineSensitizedTripletTriplet2008,
  title = {Pd({{II}}) {{Phthalocyanine-Sensitized Triplet}}-{{Triplet Annihilation}} from {{Rubrene}}},
  author = {Singh-Rachford, Tanya N. and Castellano, Felix N.},
  year = {2008},
  journal = {The Journal of Physical Chemistry A},
  shortjournal = {J. Phys. Chem. A},
  volume = {112},
  number = {16},
  pages = {3550--3556},
  publisher = {American Chemical Society},
  issn = {1089-5639},
  doi = {10.1021/jp7111878},
  url = {https://doi.org/10.1021/jp7111878},
  urlaccess = {2025-10-06}
}

@article{yongEntangledTripletPair2017,
  title = {The Entangled Triplet Pair State in Acene and Heteroacene Materials},
  author = {Yong, Chaw Keong and Musser, Andrew J. and Bayliss, Sam L. and Lukman, Steven and Tamura, Hiroyuki and Bubnova, Olga and Hallani, Rawad K. and Meneau, Aurélie and Resel, Roland and Maruyama, Munetaka and Hotta, Shu and Herz, Laura M. and Beljonne, David and Anthony, John E. and Clark, Jenny and Sirringhaus, Henning},
  year = {2017},
  journal = {Nature Communications},
  shortjournal = {Nat Commun},
  volume = {8},
  number = {1},
  pages = {15953},
  publisher = {Nature Publishing Group},
  issn = {2041-1723},
  doi = {10.1038/ncomms15953},
  url = {https://www.nature.com/articles/ncomms15953},
  urlaccess = {2025-10-06},
  langid = {english}
}

@article{bossanyiSpinStatisticsTriplet2021,
  title = {Spin {{Statistics}} for {{Triplet}}–{{Triplet Annihilation Upconversion}}: {{Exchange Coupling}}, {{Intermolecular Orientation}}, and {{Reverse Intersystem Crossing}}},
  shorttitle = {Spin {{Statistics}} for {{Triplet}}–{{Triplet Annihilation Upconversion}}},
  author = {Bossanyi, David G. and Sasaki, Yoichi and Wang, Shuangqing and Chekulaev, Dimitri and Kimizuka, Nobuo and Yanai, Nobuhiro and Clark, Jenny},
  year = {2021},
  journal = {JACS Au},
  shortjournal = {JACS Au},
  volume = {1},
  number = {12},
  pages = {2188--2201},
  publisher = {American Chemical Society},
  doi = {10.1021/jacsau.1c00322},
  url = {https://doi.org/10.1021/jacsau.1c00322},
  urlaccess = {2025-10-06}
}

@article{radiunasTripletSingletExciton2022a,
  title = {Triplet and Singlet Exciton Diffusion in Disordered Rubrene Films: Implications for Photon Upconversion},
  shorttitle = {Triplet and Singlet Exciton Diffusion in Disordered Rubrene Films},
  author = {Radiunas, Edvinas and Dapkevičius, Manvydas and Raišys, Steponas and Kazlauskas, Karolis},
  year = {2022},
  journal = {Physical Chemistry Chemical Physics},
  shortjournal = {Phys. Chem. Chem. Phys.},
  volume = {24},
  number = {39},
  pages = {24345--24352},
  publisher = {The Royal Society of Chemistry},
  issn = {1463-9084},
  doi = {10.1039/D2CP02798D},
  url = {https://pubs.rsc.org/en/content/articlelanding/2022/cp/d2cp02798d},
  urlaccess = {2025-10-28},
  langid = {english}
}

@article{wuNanocrystalSensitizedInfraredtoVisibleUpconversion2021,
  title = {Nanocrystal-{{Sensitized Infrared-to-Visible Upconversion}} in a {{Microcavity}} under {{Subsolar Flux}}},
  author = {Wu, Mengfei and Lin, Ting-An and Tiepelt, Jan O. and Bulović, Vladimir and Baldo, Marc A.},
  year = {2021},
  journal = {Nano Letters},
  shortjournal = {Nano Lett.},
  volume = {21},
  number = {2},
  pages = {1011--1016},
  publisher = {American Chemical Society},
  issn = {1530-6984},
  doi = {10.1021/acs.nanolett.0c04060},
  url = {https://doi.org/10.1021/acs.nanolett.0c04060},
  urlaccess = {2025-11-21}
}

@article{kytkaRealtimeObservationOxidation2007,
  title = {Real-Time Observation of Oxidation and Photo-Oxidation of Rubrene Thin Films by Spectroscopic Ellipsometry},
  author = {Kytka, M. and Gerlach, A. and Schreiber, F. and Kováč, J.},
  year = {2007},
  journal = {Applied Physics Letters},
  shortjournal = {Appl. Phys. Lett.},
  volume = {90},
  number = {13},
  pages = {131911},
  issn = {0003-6951},
  doi = {10.1063/1.2717561},
  url = {https://doi.org/10.1063/1.2717561},
  urlaccess = {2025-12-03}
}

@article{polakManipulatingMoleculesStrong2020,
  title = {Manipulating Molecules with Strong Coupling: Harvesting Triplet Excitons in Organic Exciton Microcavities},
  shorttitle = {Manipulating Molecules with Strong Coupling},
  author = {Polak, Daniel and Jayaprakash, Rahul and Lyons, Thomas P. and Martínez-Martínez, Luis Á and Leventis, Anastasia and Fallon, Kealan J. and Coulthard, Harriet and Bossanyi, David G. and Georgiou, Kyriacos and Anthony J. Petty, I. I. and Anthony, John and Bronstein, Hugo and Yuen-Zhou, Joel and Tartakovskii, Alexander I. and Clark, Jenny and Musser, Andrew J.},
  year = {2020},
  journal = {Chemical Science},
  shortjournal = {Chem. Sci.},
  volume = {11},
  number = {2},
  pages = {343--354},
  publisher = {The Royal Society of Chemistry},
  issn = {2041-6539},
  doi = {10.1039/C9SC04950A},
  url = {https://pubs.rsc.org/en/content/articlelanding/2020/sc/c9sc04950a},
  urlaccess = {2026-01-29},
  langid = {english}
}

@article{bucchieriTripletTripletAnnihilation2025,
  title = {Triplet–Triplet Annihilation Upconversion Sensitized with Nanocrystals for a New Generation of Photocatalytic Systems},
  author = {Bucchieri, Michele and S.~Freyria, Francesca and Bonelli, Barbara},
  year = {2025},
  journal = {Journal of Materials Chemistry A},
  volume = {13},
  number = {24},
  pages = {18115--18145},
  publisher = {Royal Society of Chemistry},
  doi = {10.1039/D5TA01541C},
  url = {https://pubs.rsc.org/en/content/articlelanding/2025/ta/d5ta01541c},
  urlaccess = {2026-01-30},
  langid = {english}
}

@article{sasakiNearInfraredOptogeneticGenome2019,
  title = {Near-{{Infrared Optogenetic Genome Engineering Based}} on {{Photon-Upconversion Hydrogels}}},
  author = {Sasaki, Yoichi and Oshikawa, Mio and Bharmoria, Pankaj and Kouno, Hironori and Hayashi-Takagi, Akiko and Sato, Moritoshi and Ajioka, Itsuki and Yanai, Nobuhiro and Kimizuka, Nobuo},
  year = {2019},
  journal = {Angewandte Chemie International Edition},
  volume = {58},
  number = {49},
  pages = {17827--17833},
  issn = {1521-3773},
  doi = {10.1002/anie.201911025},
  url = {https://onlinelibrary.wiley.com/doi/abs/10.1002/anie.201911025},
  urlaccess = {2026-01-30},
  langid = {english}
}

@article{wangMolecularNearinfraredTriplettriplet2024,
  title = {Molecular Near-Infrared Triplet-Triplet Annihilation Upconversion with Eigen Oxygen Immunity},
  author = {Wang, Xinyu and Ding, Fangwei and Jia, Tao and Li, Feng and Ding, Xiping and Deng, Ruibin and Lin, Kaifeng and Yang, Yulin and Wu, Wenzhi and Xia, Debin and Chen, Guanying},
  year = {2024},
  journal = {Nature Communications},
  shortjournal = {Nat Commun},
  volume = {15},
  number = {1},
  pages = {2157},
  publisher = {Nature Publishing Group},
  issn = {2041-1723},
  doi = {10.1038/s41467-024-46541-z},
  url = {https://www.nature.com/articles/s41467-024-46541-z},
  urlaccess = {2026-01-30},
  langid = {english}
}

@article{carrodRecentAdvancesTriplet2022,
  title = {Recent Advances in Triplet–Triplet Annihilation Upconversion and Singlet Fission, towards Solar Energy Applications},
  author = {Carrod, Andrew J. and Gray, Victor and Börjesson, Karl},
  year = {2022},
  journal = {Energy \& Environmental Science},
  shortjournal = {Energy Environ. Sci.},
  volume = {15},
  number = {12},
  pages = {4982--5016},
  publisher = {The Royal Society of Chemistry},
  issn = {1754-5706},
  doi = {10.1039/D2EE01600A},
  url = {https://pubs.rsc.org/en/content/articlelanding/2022/ee/d2ee01600a},
  urlaccess = {2026-01-30},
  langid = {english}
}

@article{ekins-daukesMolecularApproachIntermediate2008,
  title = {A Molecular Approach to the Intermediate Band Solar Cell: {{The}} Symmetric Case},
  shorttitle = {A Molecular Approach to the Intermediate Band Solar Cell},
  author = {Ekins-Daukes, N. J. and Schmidt, T. W.},
  year = {2008},
  journal = {Applied Physics Letters},
  shortjournal = {Appl. Phys. Lett.},
  volume = {93},
  number = {6},
  pages = {063507},
  issn = {0003-6951},
  doi = {10.1063/1.2970157},
  url = {https://doi.org/10.1063/1.2970157},
  urlaccess = {2026-01-30}
}

@article{goldschmidtUpconversionPhotovoltaicsReview2015,
  title = {Upconversion for {{Photovoltaics}} – a {{Review}} of {{Materials}}, {{Devices}} and {{Concepts}} for {{Performance Enhancement}}},
  author = {Goldschmidt, Jan Christoph and Fischer, Stefan},
  year = {2015},
  journal = {Advanced Optical Materials},
  volume = {3},
  number = {4},
  pages = {510--535},
  issn = {2195-1071},
  doi = {10.1002/adom.201500024},
  url = {https://onlinelibrary.wiley.com/doi/abs/10.1002/adom.201500024},
  urlaccess = {2026-01-30},
  langid = {english}
}

@article{huangTripletTripletAnnihilation2024,
  title = {Triplet–Triplet Annihilation Photon Upconversion-Mediated Photochemical Reactions},
  author = {Huang, Ling and Han, Gang},
  year = {2024-04},
  journal = {Nature Reviews Chemistry},
  shortjournal = {Nat Rev Chem},
  volume = {8},
  number = {4},
  pages = {238--255},
  publisher = {Nature Publishing Group},
  issn = {2397-3358},
  doi = {10.1038/s41570-024-00585-3},
  url = {https://www.nature.com/articles/s41570-024-00585-3},
  urlaccess = {2026-01-30},
  langid = {english}
}

@article{kinoshitaPhotonUpconvertingSolid2020a,
  title = {Photon {{Upconverting Solid Films}} with {{Improved Efficiency}} for {{Endowing Perovskite Solar Cells}} with {{Near-Infrared Sensitivity}}},
  author = {Kinoshita, Mika and Sasaki, Yoichi and Amemori, Shogo and Harada, Naoyuki and Hu, Zhanhao and Liu, Zonghao and Ono, Luis K. and Qi, Yabing and Yanai, Nobuhiro and Kimizuka, Nobuo},
  year = {2020},
  journal = {ChemPhotoChem},
  volume = {4},
  number = {11},
  pages = {5271--5278},
  issn = {2367-0932},
  doi = {10.1002/cptc.202000143},
  url = {https://onlinelibrary.wiley.com/doi/abs/10.1002/cptc.202000143},
  urlaccess = {2026-01-30},
  langid = {english}
}

@article{barnesFluorescenceInterfacesRole1998,
  title = {Fluorescence near Interfaces: {{The}} Role of Photonic Mode Density},
  shorttitle = {Fluorescence near Interfaces},
  author = {Barnes, W. L.},
  year = {1998},
  journal = {Journal of Modern Optics},
  volume = {45},
  number = {4},
  pages = {661--699},
  publisher = {Taylor \& Francis},
  issn = {0950-0340},
  doi = {10.1080/09500349808230614},
  url = {https://doi.org/10.1080/09500349808230614},
  urlaccess = {2026-01-31}
}

@article{benistyImpactPlanarMicrocavity1998,
  title = {Impact of Planar Microcavity Effects on Light Extraction-{{Part I}}: Basic Concepts and Analytical Trends},
  shorttitle = {Impact of Planar Microcavity Effects on Light Extraction-{{Part I}}},
  author = {Benisty, H. and De Neve, H. and Weisbuch, C.},
  year = {1998-09},
  journal = {IEEE Journal of Quantum Electronics},
  volume = {34},
  number = {9},
  pages = {1612--1631},
  issn = {1558-1713},
  doi = {10.1109/3.709578},
  url = {https://ieeexplore.ieee.org/document/709578},
  urlaccess = {2026-01-31}
}

@article{linExaminingMicrocavityOrganic2005,
  title = {Examining Microcavity Organic Light-Emitting Devices Having Two Metal Mirrors},
  author = {Lin, Chun-Liang and Lin, Hao-Wu and Wu, Chung-Chih},
  year = {2005},
  journal = {Applied Physics Letters},
  shortjournal = {Appl. Phys. Lett.},
  volume = {87},
  number = {2},
  pages = {021101},
  issn = {0003-6951},
  doi = {10.1063/1.1988985},
  url = {https://doi.org/10.1063/1.1988985},
  urlaccess = {2026-01-31}
}

@article{nowyLightExtractionOptical2008,
  title = {Light Extraction and Optical Loss Mechanisms in Organic Light-Emitting Diodes: {{Influence}} of the Emitter Quantum Efficiency},
  shorttitle = {Light Extraction and Optical Loss Mechanisms in Organic Light-Emitting Diodes},
  author = {Nowy, Stefan and Krummacher, Benjamin C. and Frischeisen, Jörg and Reinke, Nils A. and Brütting, Wolfgang},
  year = {2008},
  journal = {Journal of Applied Physics},
  shortjournal = {J. Appl. Phys.},
  volume = {104},
  number = {12},
  pages = {123109},
  issn = {0021-8979},
  doi = {10.1063/1.3043800},
  url = {https://doi.org/10.1063/1.3043800},
  urlaccess = {2026-01-31}
}

@article{radiunasNIRtovisPhotonUpconversion2021a,
  title = {{{NIR-to-vis}} Photon Upconversion in Rubrenes with Increasing Structural Complexity},
  author = {Radiunas, Edvinas and Dapkevičius, Manvydas and Naimovičius, Lukas and Baronas, Paulius and Raišys, Steponas and Juršėnas, Saulius and Jozeliūnaitė, Augustina and Javorskis, Tomas and Šinkevičiūtė, Ugnė and Orentas, Edvinas and Kazlauskas, Karolis},
  year = {2021},
  journal = {Journal of Materials Chemistry C},
  shortjournal = {J. Mater. Chem. C},
  volume = {9},
  number = {12},
  pages = {4359--4366},
  publisher = {The Royal Society of Chemistry},
  issn = {2050-7534},
  doi = {10.1039/D1TC00296A},
  url = {https://pubs.rsc.org/en/content/articlelanding/2021/tc/d1tc00296a},
  urlaccess = {2026-01-31},
  langid = {english}
}

@article{irkhinDefectDensityDependent2016,
  title = {Defect Density Dependent Photoluminescence Yield and Triplet Diffusion Length in Rubrene},
  author = {Irkhin, Pavel and Biaggio, Ivan and Zimmerling, Tino and Döbeli, Max and Batlogg, Bertram},
  year = {2016},
  journal = {Applied Physics Letters},
  shortjournal = {Appl. Phys. Lett.},
  volume = {108},
  number = {6},
  pages = {063302},
  issn = {0003-6951},
  doi = {10.1063/1.4941756},
  url = {https://doi.org/10.1063/1.4941756},
  urlaccess = {2026-02-11}
}

@article{jordanEfficiencyEnhancementMicrocavity1996,
  title = {Efficiency Enhancement of Microcavity Organic Light Emitting Diodes},
  author = {Jordan, R. H. and Dodabalapur, A. and Slusher, R. E.},
  year = {1996},
  journal = {Applied Physics Letters},
  shortjournal = {Appl. Phys. Lett.},
  volume = {69},
  number = {14},
  pages = {1997--1999},
  issn = {0003-6951},
  doi = {10.1063/1.116858},
  url = {https://doi.org/10.1063/1.116858},
  urlaccess = {2026-02-11}
}

@article{radiunasCNTuningPathwaySuppress2025,
  title = {{{CN-Tuning}}: {{A Pathway}} to {{Suppress Singlet Fission}} and {{Amplify Triplet-Triplet Annihilation Upconversion}} in {{Rubrene}}},
  shorttitle = {{{CN-Tuning}}},
  author = {Radiunas, Edvinas and Naimovičius, Lukas and Baronas, Paulius and Jozeliūnaitė, Augustina and Orentas, Edvinas and Kazlauskas, Karolis},
  year = {2025},
  journal = {Advanced Optical Materials},
  volume = {13},
  number = {12},
  pages = {2403032},
  issn = {2195-1071},
  doi = {10.1002/adom.202403032},
  url = {https://onlinelibrary.wiley.com/doi/abs/10.1002/adom.202403032},
  urlaccess = {2026-02-11},
  langid = {english}
}

@article{shengTremendouslyEnhancedPhotocurrent2021,
  title = {Tremendously Enhanced Photocurrent Enabled by Triplet–Triplet Annihilation up-Conversion for High-Performance Perovskite Solar Cells},
  author = {Sheng, Wangping and Yang, Jia and Li, Xiang and Liu, Gengling and Lin, Zhuojia and Long, Juan and Xiao, Shuqin and Tan, Licheng and Chen, Yiwang},
  year = {2021},
  journal = {Energy \& Environmental Science},
  shortjournal = {Energy Environ. Sci.},
  volume = {14},
  number = {6},
  pages = {3532--3541},
  publisher = {The Royal Society of Chemistry},
  issn = {1754-5706},
  doi = {10.1039/D1EE00631B},
  url = {https://pubs.rsc.org/en/content/articlelanding/2021/ee/d1ee00631b},
  urlaccess = {2026-02-11},
  langid = {english}
}

@article{wiegholdDisorderBeneficialPerovskiteSensitized2020a,
  title = {Is {{Disorder Beneficial}} in {{Perovskite-Sensitized Solid-State Upconversion}}? {{The Role}} of {{DBP Doping}} in {{Rubrene}}},
  shorttitle = {Is {{Disorder Beneficial}} in {{Perovskite-Sensitized Solid-State Upconversion}}?},
  author = {Wieghold, Sarah and Bieber, Alexander S. and VanOrman, Zachary A. and Rodriguez, Arianna and Nienhaus, Lea},
  year = {2020},
  journal = {The Journal of Physical Chemistry C},
  shortjournal = {J. Phys. Chem. C},
  volume = {124},
  number = {33},
  pages = {18132--18140},
  publisher = {American Chemical Society},
  issn = {1932-7447},
  doi = {10.1021/acs.jpcc.0c05290},
  url = {https://doi.org/10.1021/acs.jpcc.0c05290},
  urlaccess = {2026-02-11}
}

@incollection{purcellSpontaneousEmissionProbabilities1995,
  title = {Spontaneous {{Emission Probabilities}} at {{Radio Frequencies}}},
  booktitle = {Confined {{Electrons}} and {{Photons}}: {{New Physics}} and {{Applications}}},
  author = {Purcell, E. M.},
  editor = {Burstein, Elias and Weisbuch, Claude},
  year = {1995},
  pages = {839--839},
  publisher = {Springer US},
  location = {Boston, MA},
  doi = {10.1007/978-1-4615-1963-8_40},
  url = {https://doi.org/10.1007/978-1-4615-1963-8_40},
  urlaccess = {2026-02-11},
  isbn = {978-1-4615-1963-8},
  langid = {english}
}

@book{diracPrinciplesQuantumMechanics1981,
  title = {The {{Principles}} of {{Quantum Mechanics}}},
  author = {Dirac, Paul Adrien Maurice},
  year = {1981},
  eprint = {XehUpGiM6FIC},
  eprinttype = {googlebooks},
  publisher = {Clarendon Press},
  isbn = {978-0-19-852011-5},
  langid = {english},
  pagetotal = {340}
}

@book{fermiNuclearPhysicsCourse1950,
  title = {Nuclear {{Physics}}: {{A Course Given}} by {{Enrico Fermi}} at the {{University}} of {{Chicago}}},
  shorttitle = {Nuclear {{Physics}}},
  author = {Fermi, Enrico},
  year = {1950},
  eprint = {WQtkYCWTcicC},
  eprinttype = {googlebooks},
  publisher = {University of Chicago Press},
  isbn = {978-0-226-24365-8},
  langid = {english},
  pagetotal = {266}
}

@article{visserWhoseGoldenRule2009,
  title = {Whose Golden Rule Is It Anyway?},
  author = {Visser, Taco D.},
  year = {2009},
  journal = {American Journal of Physics},
  shortjournal = {Am. J. Phys.},
  volume = {77},
  number = {6},
  pages = {487},
  issn = {0002-9505},
  doi = {10.1119/1.3096649},
  url = {https://doi.org/10.1119/1.3096649},
  urlaccess = {2026-02-11}
}

@article{irkhinAbsorptionPhotoluminescenceSpectroscopy2012,
  title = {Absorption and Photoluminescence Spectroscopy of Rubrene Single Crystals},
  author = {Irkhin, Pavel and Ryasnyanskiy, Aleksandr and Koehler, Marlus and Biaggio, Ivan},
  year = {2012},
  journal = {Physical Review B},
  shortjournal = {Phys. Rev. B},
  volume = {86},
  number = {8},
  pages = {085143},
  publisher = {American Physical Society},
  doi = {10.1103/PhysRevB.86.085143},
  url = {https://link.aps.org/doi/10.1103/PhysRevB.86.085143},
  urlaccess = {2026-02-13}
}

@article{chePhotophysicalPropertiesPalladium2018,
  title = {Photophysical Properties of Palladium/Platinum Tetrasulfonyl Phthalocyanines and Their Application in Triplet–Triplet Annihilation Upconversion},
  author = {Che, Yuanyuan and Yang, Wenbo and Tang, Geliang and Dumoulin, Fabienne and Zhao, Jianzhang and Liu, Lang and {\.{I}}\c{s}ci, Umit},
  year = {2018},
  journal = {Journal of Materials Chemistry C},
  shortjournal = {J. Mater. Chem. C},
  volume = {6},
  number = {21},
  pages = {5785--5793},
  publisher = {The Royal Society of Chemistry},
  issn = {2050-7534},
  doi = {10.1039/C8TC00513C},
  url = {https://pubs.rsc.org/en/content/articlelanding/2018/tc/c8tc00513c},
  urlaccess = {2026-02-25},
  langid = {english}
}

@article{hanFirstTransitionMetal2017,
  title = {The First Transition Metal Phthalocyanines: Sensitizing Rubrene Emission Based on Triplet–Triplet Annihilation},
  shorttitle = {The First Transition Metal Phthalocyanines},
  author = {Han, Jianlei and Zhang, Fei and You, Jing and Hiroaki, Yonemura and Yamada, Sunao and Morifuji, Toru and Wang, Shirong and Li, Xianggao},
  year = {2017},
  journal = {Photochemical \& Photobiological Sciences},
  shortjournal = {Photochem Photobiol Sci},
  volume = {16},
  number = {9},
  pages = {1384--1390},
  issn = {1474-9092},
  doi = {10.1039/c6pp00464d},
  url = {https://doi.org/10.1039/c6pp00464d},
  urlaccess = {2026-02-25},
  langid = {english}
}

@article{linStrategiesHighPerformanceSolidState2020a,
  title = {Strategies for {{High-Performance Solid-State Triplet}}–{{Triplet-Annihilation-Based Photon Upconversion}}},
  author = {Lin, Ting-An and Perkinson, Collin F. and Baldo, Marc A.},
  year = {2020},
  journal = {Advanced Materials},
  volume = {32},
  number = {26},
  pages = {1908175},
  issn = {1521-4095},
  doi = {10.1002/adma.201908175},
  url = {https://onlinelibrary.wiley.com/doi/abs/10.1002/adma.201908175},
  urlaccess = {2026-02-25},
  langid = {english}
}

@article{lissauNearInfraredVisiblePhoton2021a,
  title = {Near-{{Infrared}} to {{Visible Photon Upconversion}} by {{Palladium}}({{II}}) {{Octabutoxyphthalocyanine}} and {{Rubrene}} in the {{Solid State}}},
  author = {Lissau, Jonas Sandby and Khelfallah, Malika and Madsen, Morten},
  year = {2021},
  journal = {The Journal of Physical Chemistry C},
  shortjournal = {J. Phys. Chem. C},
  volume = {125},
  number = {46},
  pages = {25643--25650},
  publisher = {American Chemical Society},
  issn = {1932-7447},
  doi = {10.1021/acs.jpcc.1c06945},
  url = {https://doi.org/10.1021/acs.jpcc.1c06945},
  urlaccess = {2026-02-25}
}

@article{maPhaseShiftPenetration2007,
  title = {Phase Shift and Penetration Depth of Metal Mirrors in a Microcavity Structure},
  author = {Ma, Fengying and Liu, Xingyuan},
  year = {2007},
  journal = {Applied Optics},
  shortjournal = {Appl. Opt., AO},
  volume = {46},
  number = {25},
  pages = {6247--6250},
  publisher = {Optica Publishing Group},
  issn = {2155-3165},
  doi = {10.1364/AO.46.006247},
  url = {https://opg.optica.org/ao/abstract.cfm?uri=ao-46-25-6247},
  urlaccess = {2026-02-25},
  langid = {english}
}

@article{murakamiKineticsPhotonUpconversion2021,
  title = {Kinetics of Photon Upconversion by Triplet–Triplet Annihilation: A Comprehensive Tutorial},
  shorttitle = {Kinetics of Photon Upconversion by Triplet–Triplet Annihilation},
  author = {Murakami, Yoichi and Kamada, Kenji},
  year = {2021},
  journal = {Physical Chemistry Chemical Physics},
  shortjournal = {Phys. Chem. Chem. Phys.},
  volume = {23},
  number = {34},
  pages = {18268--18282},
  publisher = {The Royal Society of Chemistry},
  issn = {1463-9084},
  doi = {10.1039/D1CP02654B},
  url = {https://pubs.rsc.org/en/content/articlelanding/2021/cp/d1cp02654b},
  urlaccess = {2026-02-25},
  langid = {english}
}

@article{radiunasEfficientNIRtovisPhoton2022a,
  title = {Efficient {{NIR-to-vis}} Photon Upconversion in Binary Rubrene Films Deposited by Simplified Thermal Evaporation},
  author = {Radiunas, Edvinas and Naimovičius, Lukas and Raišys, Steponas and Jozeliūnaitė, Augustina and Orentas, Edvinas and Kazlauskas, Karolis},
  year = {2022},
  journal = {Journal of Materials Chemistry C},
  shortjournal = {J. Mater. Chem. C},
  volume = {10},
  number = {16},
  pages = {6314--6322},
  publisher = {The Royal Society of Chemistry},
  issn = {2050-7534},
  doi = {10.1039/D1TC05332A},
  url = {https://pubs.rsc.org/en/content/articlelanding/2022/tc/d1tc05332a},
  urlaccess = {2026-02-25},
  langid = {english}
}

@article{radiunasImpactTbutylSubstitution2020a,
  title = {Impact of T-Butyl Substitution in a Rubrene Emitter for Solid State {{NIR-to-visible}} Photon Upconversion},
  author = {Radiunas, Edvinas and Dapkevičius, Manvydas and Raišys, Steponas and Juršėnas, Saulius and Jozeliūnaitė, Augustina and Javorskis, Tomas and Šinkevičiūtė, Ugnė and Orentas, Edvinas and Kazlauskas, Karolis},
  year = {2020},
  journal = {Physical Chemistry Chemical Physics},
  shortjournal = {Phys. Chem. Chem. Phys.},
  volume = {22},
  number = {14},
  pages = {7392--7403},
  publisher = {The Royal Society of Chemistry},
  issn = {1463-9084},
  doi = {10.1039/D0CP00144A},
  url = {https://pubs.rsc.org/en/content/articlelanding/2020/cp/d0cp00144a},
  urlaccess = {2026-02-25},
  langid = {english}
}

@article{radiunasUnderstandingLimitationsNIRtovisible2020a,
  title = {Understanding the Limitations of {{NIR-to-visible}} Photon Upconversion in Phthalocyanine-Sensitized Rubrene Systems},
  author = {Radiunas, Edvinas and Raišys, Steponas and Juršėnas, Saulius and Jozeliūnaitė, Augustina and Javorskis, Tomas and Šinkevičiūtė, Ugnė and Orentas, Edvinas and Kazlauskas, Karolis},
  year = {2020},
  journal = {Journal of Materials Chemistry C},
  shortjournal = {J. Mater. Chem. C},
  volume = {8},
  number = {16},
  pages = {5525--5534},
  publisher = {The Royal Society of Chemistry},
  issn = {2050-7534},
  doi = {10.1039/C9TC06031F},
  url = {https://pubs.rsc.org/en/content/articlelanding/2020/tc/c9tc06031f},
  urlaccess = {2026-02-25},
  langid = {english}
}

@article{sawaEnhancingNIRtovisiblePhoton2023a,
  title = {Enhancing {{NIR-to-visible}} Photon Upconversion in Cast Solid by Introducing Bulky Substituents in Rubrene and by Suppressing Back Energy Transfer},
  author = {Sawa, Akane and Shimada, Shota and Tripathi, Neeti and Heck, Claire and Tachibana, Hiroaki and Koyama, Emiko and Mizokuro, Toshiko and Hirao, Yasukazu and Kubo, Takashi and Tamai, Naoto and Kuzuhara, Daiki and Yamada, Hiroko and Kamada, Kenji},
  year = {2023},
  journal = {Journal of Materials Chemistry C},
  shortjournal = {J. Mater. Chem. C},
  volume = {11},
  number = {25},
  pages = {8502--8513},
  publisher = {The Royal Society of Chemistry},
  issn = {2050-7534},
  doi = {10.1039/D3TC00853C},
  url = {https://pubs.rsc.org/en/content/articlelanding/2023/tc/d3tc00853c},
  urlaccess = {2026-02-25},
  langid = {english}
}

@article{wangHighlyefficientRedtoyellowGreen2021a,
  title = {Highly-Efficient Red-to-Yellow/Green Upconversion Sensitized by Phthalocyanine Palladium with High Triplet Energy-Level},
  author = {Wang, Kai and Huang, Suqin and Ding, Ping and Liang, Zuoqin and Chen, Shuoran and Li, Lin and Ye, Changqing and Wang, Xiaomei},
  year = {2021},
  journal = {RSC Advances},
  shortjournal = {RSC Adv.},
  volume = {11},
  number = {29},
  pages = {17755--17759},
  publisher = {The Royal Society of Chemistry},
  issn = {2046-2069},
  doi = {10.1039/D1RA02528G},
  url = {https://pubs.rsc.org/en/content/articlelanding/2021/ra/d1ra02528g},
  urlaccess = {2026-02-25},
  langid = {english}
}

@article{duanEfficientSolidstateInfraredtovisible2022a,
  title = {Efficient Solid-State Infrared-to-Visible Photon Upconversion on Atomically Thin Monolayer Semiconductors},
  author = {Duan, Jiaru and Liu, Yanping and Zhang, Yongqing and Chen, Zeng and Xu, Xuehui and Ye, Lei and Wang, Zukun and Yang, Yang and Zhang, Delong and Zhu, Haiming},
  year = {2022},
  journal = {Science Advances},
  volume = {8},
  number = {43},
  pages = {eabq4935},
  publisher = {American Association for the Advancement of Science},
  doi = {10.1126/sciadv.abq4935},
  url = {https://www.science.org/doi/10.1126/sciadv.abq4935},
  urlaccess = {2026-02-26}
}

@article{jonesUpconversionQuantumYield2021,
  title = {The Upconversion Quantum Yield ({{UCQY}}): A Review to Standardize the Measurement Methodology, Improve Comparability, and Define Efficiency Standards},
  shorttitle = {The Upconversion Quantum Yield ({{UCQY}})},
  author = {Jones, Callum M. S. and Gakamsky, Anna and Marques-Hueso, Jose},
  year = {2021},
  journal = {Science and Technology of Advanced Materials},
  volume = {22},
  number = {1},
  eprint = {34992499},
  eprinttype = {pubmed},
  pages = {810--848},
  publisher = {Taylor \& Francis},
  issn = {1468-6996},
  doi = {10.1080/14686996.2021.1967698},
  url = {https://doi.org/10.1080/14686996.2021.1967698},
  urlaccess = {2026-02-26}
}

@article{narayananAlleviatingParasiticBack2025,
  title = {Alleviating {{Parasitic Back Energy Transfer Enhances Thin Film Upconversion}}},
  author = {Narayanan, Pournima and Hu, Manchen and Pucurimay, Linda and Gallegos, Arynn O. and Zhou, Qi and Belliveau, Emma and Ahmed, Ghada H. and Fernández, Sebastian and Michaels, William and Murrietta, Natalia and Mutatu, Vongaishe E. and Feng, Demeng and Hamid, Rabeeya and Yap, Kyra M. K. and Schloemer, Tracy H. and Jaramillo, Thomas F. and Kats, Mikhail A. and Congreve, Daniel N.},
  year = {2025},
  journal = {Advanced Optical Materials},
  volume = {13},
  number = {17},
  pages = {2500252},
  issn = {2195-1071},
  doi = {10.1002/adom.202500252},
  url = {https://onlinelibrary.wiley.com/doi/abs/10.1002/adom.202500252},
  urlaccess = {2026-02-26},
  langid = {english}
}

@article{singh-rachfordPhotonUpconversionBased2010,
  title = {Photon Upconversion Based on Sensitized Triplet–Triplet Annihilation},
  author = {Singh-Rachford, Tanya N. and Castellano, Felix N.},
  year = {2010},
  journal = {Coordination Chemistry Reviews},
  shortjournal = {Coordination Chemistry Reviews},
  series = {18th {{International Symposium}} on the {{Photochemistry}} and {{Photophysics}} of {{Coordination Compounds Sapporo}}, 2009},
  volume = {254},
  number = {21},
  pages = {2560--2573},
  issn = {0010-8545},
  doi = {10.1016/j.ccr.2010.01.003},
  url = {https://www.sciencedirect.com/science/article/pii/S0010854510000093},
  urlaccess = {2026-02-26}
}

@article{torrePhthalocyaninesOldDyes2007,
  title = {Phthalocyanines: Old Dyes, New Materials. {{Putting}} Color in Nanotechnology},
  shorttitle = {Phthalocyanines},
  author = {de la Torre, Gema and Claessens, Christian G. and Torres, Tom{\'{a}}s},
  year = {2007},
  journal = {Chemical Communications},
  shortjournal = {Chem. Commun.},
  number = {20},
  pages = {2000--2015},
  publisher = {The Royal Society of Chemistry},
  issn = {1364-548X},
  doi = {10.1039/B614234F},
  url = {https://pubs.rsc.org/en/content/articlelanding/2007/cc/b614234f},
  urldate = {2026-02-26},
  langid = {english}
}

@article{xiaAchievingSpintripletExciton2020,
  title = {Achieving Spin-Triplet Exciton Transfer between Silicon and Molecular Acceptors for Photon Upconversion},
  author = {Xia, Pan and Raulerson, Emily K. and Coleman, Devin and Gerke, Carter S. and Mangolini, Lorenzo and Tang, Ming Lee and Roberts, Sean T.},
  year = {2020-02},
  journal = {Nature Chemistry},
  shortjournal = {Nat. Chem.},
  volume = {12},
  number = {2},
  pages = {137--144},
  publisher = {Nature Publishing Group},
  issn = {1755-4349},
  doi = {10.1038/s41557-019-0385-8},
  url = {https://www.nature.com/articles/s41557-019-0385-8},
  urlaccess = {2026-02-26},
  langid = {english}
}

@article{nicklinMINERVAFacilityStudy2017,
  title = {{{MINERVA}}: {{A}} Facility to Study {{Microstructure}} and {{INterface Evolution}} in {{Realtime}} under {{VAcuum}}},
  shorttitle = {{{MINERVA}}},
  author = {Nicklin, Chris and Martinez-Hardigree, Josue and Warne, Adam and Green, Stephen and Burt, Martin and Naylor, John and Dorman, Adam and Wicks, Dean and Din, Salahud and Riede, Moritz},
  year = {2017},
  journal = {Review of Scientific Instruments},
  shortjournal = {Rev. Sci. Instrum.},
  volume = {88},
  number = {10},
  pages = {103901},
  issn = {0034-6748},
  doi = {10.1063/1.4989761},
  url = {https://doi.org/10.1063/1.4989761},
  urlaccess = {2026-03-09}
}

@article{ashiotisFastAzimuthalIntegration2015,
  title = {The Fast Azimuthal Integration {{Python}} Library: {{pyFAI}}},
  shorttitle = {The Fast Azimuthal Integration {{Python}} Library},
  author = {Ashiotis, Giannis and Deschildre, Aurore and Nawaz, Zubair and Wright, Jonathan P. and Karkoulis, Dimitrios and Picca, Frédéric Emmanuel and Kieffer, Jérôme},
  year = {2015},
  journal = {Journal of Applied Crystallography},
  shortjournal = {J Appl Crystallogr},
  volume = {48},
  eprint = {25844080},
  eprinttype = {pubmed},
  pages = {510--519},
  issn = {0021-8898},
  doi = {10.1107/S1600576715004306},
  issue = {Pt 2},
  langid = {english},
  pmcid = {PMC4379438}
}

@article{jiSemiconductorPlasmonEnhanced2019,
  title = {Semiconductor Plasmon Enhanced Monolayer Upconversion Nanoparticles for High Performance Narrowband Near-Infrared Photodetection},
  author = {Ji, Yanan and Xu, Wen and Li, Dongyu and Zhou, Donglei and Chen, Xu and Ding, Nan and Li, Jing and Wang, Nan and Bai, Xue and Song, Hongwei},
  year = {2019},
  journal = {Nano Energy},
  shortjournal = {Nano Energy},
  volume = {61},
  pages = {211--220},
  issn = {2211-2855},
  doi = {10.1016/j.nanoen.2019.04.064},
  url = {https://www.sciencedirect.com/science/article/pii/S2211285519303659},
  urlaccess = {2026-03-16}
}

@article{limbergTripletTripletAnnihilation2022,
  title = {Triplet--{{Triplet Annihilation Photopolymerization}} for {{High}}-{{Resolution 3D Printing}}},
  author = {Limberg, David K. and Kang, Ji-Hwan and Hayward, Ryan C.},
  year = {2022},
  journal = {Journal of the American Chemical Society},
  volume = {144},
  number = {12},
  pages = {5226--5232},
  issn = {0002-7863},
  doi = {10.1021/jacs.1c11022},
  url = {https://doi.org/10.1021/jacs.1c11022},
  urlaccess = {2026-03-16},
  langid = {english}
}

@article{yeKineticsEfficiencyTripletTriplet2021,
  title = {Kinetics and {{Efficiency}} of {{Triplet}}-{{Triplet Annihilation Photon Upconversion}} under {{Pulsed Excitation Conditions}}},
  author = {Ye, Chen and Albinsson, Bo and B{\"o}rjesson, Karl},
  year = {2021},
  journal = {Chemistry -- Methods},
  volume = {1},
  number = {3},
  pages = {136--142},
  issn = {2628-9725},
  doi = {10.1002/cmtd.202000046},
  url = {https://chemistry-europe.onlinelibrary.wiley.com/doi/10.1002/cmtd.202000046},
  urlaccess = {2026-10-13},
  langid = {english}
}

\end{document}